\documentclass[12pt,preprint]{aastex}
\usepackage{graphicx}
\usepackage{natbib}
\begin{document}
\title{The Christiansen Effect in Saturn's narrow dusty rings
and the spectral identification of clumps in the F ring}
\author{M.M. Hedman$^{*,a}$, P.D. Nicholson$^a$, M.R. Showalter$^b$, \\
R.H. Brown$^c$, B.J. Buratti$^d$, R.N. Clark$^e$, K. Baines$^d$, C. Sotin$^d$}
\affil{$^*$ Corresponding Author, mmhedman@astro.cornell.edu
$^a$ Department of Astronomy, Cornell University, Ithaca NY 14853\\
$^b$ SETI Institute, Mountain View CA 94043 \\
$^c$ Lunar and Planetary Laboratory, University of Arizona, Tuscon AZ 85721 \\
$^d$ Jet Propulsion Laboratory, Pasadena CA 91109 \\
$^e$ USGS, Denver CO 80225  }

{Stellar occultations by Saturn's rings observed with
the Visual and Infrared  Mapping Spectrometer (VIMS) onboard the Cassini spacecraft
reveal that dusty features such as the F ring and the ringlets 
in the Encke and the Laplace Gaps have distinctive infrared transmission spectra.
These spectra show a narrow optical depth minimum at wavelengths around 
2.87 $\mu$m. This minimum is likely due to 
the Christiansen Effect, a reduction in the extinction of small  
particles when their (complex) refractive index is close to that of the surrounding medium. Simple Mie-scattering models demonstrate that the strength of this opacity dip  is sensitive to the size distribution of particles between
1 and 100 $\mu$m across.  Furthermore, the spatial resolution of the occultation data is sufficient to  reveal  variations in the transmission spectra within and among
these rings. In both the Encke Gap ringlets and F ring, the opacity dip weakens with increasing local optical depth, 
which is consistent with the larger particles being concentrated near
the cores of these rings.
The Encke Gap ringlets also show systematically weaker opacity dips
than the F ring and Laplace Gap ringlet, implying that the former has
a smaller fraction of grains less than $\sim30 \mu$m across. However, the strength 
of the opacity dip varies most dramatically within the F ring; certain compact regions of 
enhanced optical  depth lack an opacity dip and therefore 
appear to have a greatly reduced fraction of grains in the few-micron size range. 
Such spectrally-identifiable structures probably represent a subset
of the compact optically-thick clumps observed by other 
Cassini instruments.
These variations in the ring's particle size distribution can provide
new insights into the processes of grain aggregation, disruption and
transport within dusty rings. For example, 
the unusual spectral properties of the F-ring clumps 
could perhaps be ascribed to small 
grains adhering onto the surface of larger particles in
regions of anomalously low velocity dispersion.}

\maketitle

\section{Introduction}

Stellar occultations observed by the Visual
and Infrared Mapping Spectrometer (VIMS) onboard
the Cassini Spacecraft have already provided constraints
on the geometry of self-gravity wakes in the A and B rings
\citep{Hedman07,NH10} and the architecture of the Cassini Division 
\citep{Hedman10}. However, these analyses only
used a fraction of the information returned by VIMS, 
because they were  based on light curves derived
from a single spectral channel. During each
occultation, VIMS  simultaneously measures the opacity of the rings over
a range of wavelengths from 0.85 to 5.0 $\mu$m, which 
includes the strong water-ice absorption band at 3.1 $\mu$m.
Thus each stellar occultation can in principle provide
high-spatial-resolution transmission spectra of the rings.
In practice, the optical depth of most regions in Saturn's main rings does not
vary with wavelength because nearly all of the particles in the 
main rings are much larger  than the near-infrared wavelengths observed. 
In this geometrical optics limit the transmission is 
essentially independent of wavelength. 

However, the transmission can vary with wavelength when the 
particles are comparable in size to the observing wavelength.
Several features in Saturn's rings are strongly forward scattering in
the visible and near-infrared, 
indicating that they are composed primarily of micron-sized
grains \citep{Horanyi09} and this has been confirmed
for the F-ring by detailed spectrophotometric analyses 
\citep{Showalter92, Vahidinia11}. Searches for transmission variations 
using Earth-based occultations
of the F ring did not reveal any statistically significant trends at
ultra-violet or visible wavelengths
 \citep{Bosh02}, but VIMS
occultations by the F ring and other similarly dusty ringlets
in the A-ring's Encke Gap and the Cassini Division's Laplace Gap
have revealed a narrow opacity dip in the
transmission spectra near 2.87 $\mu$m.
As discussed in detail below, this feature provides novel constraints
on the composition and structure of these dusty rings. 
Of particular interest is the ability of near-infrared stellar occultations to discern
variations in the rings' particle size distribution on finer spatial scales
than otherwise possible.

Our analysis begins by describing the relevant observations and 
how they were processed to obtain light curves. Second,
we examine an illustrative example of the transmission spectra
and demonstrate how the observed feature can be explained in terms of
the Christiansen Effect associated with the strong water-ice 
absorption band centered at 3.1$\mu$m. We then
discuss how the strength of this feature relates to the local
particle size distribution. In the future,  we expect that combining 
these transmission spectra with relevant reflectance spectra and 
phase curves will place tight constraints on the size 
distribution, but such a complete photometric analysis is beyond
the scope of this paper. Instead, we turn our attention to
the variations in the transmission spectra, 
which allow us to discern trends in the particle size
distribution. We find that the strength of the opacity dip varies systematically
among the different ringlets, demonstrating that these dusty rings
do have somewhat different particle size distributions.
In particular, we explore the spectral variations
within the F ring itself, focusing on narrow regions
where the spectral feature appears to be
highly suppressed, indicating a reduced concentration
of micron-sized particles. These features
are likely a subset of the compact optical depth enhancements
identified in both Cassini images 
\citep{Murray08, Beurle10} and UVIS stellar occultations \citep{Esposito08, Meinke10}. 
Finally, we discuss possible interpretations of the observed 
spectral variations in terms of spatially-varying particle densities
and velocity dispersions within these dusty rings.
 
\section{Observations}
\label{obssec}

VIMS is most often used to produce spatially-resolved reflectance spectra of 
planetary targets. However, VIMS is a flexible instrument that can 
also operate in an occultation mode~\citep{Brown04}. In this 
mode, the imaging capabilities are disabled, the  short-wavelength 
VIS channel of the instrument is turned off and the IR 
channel obtains a  series of spectra from a single pixel 
targeted at a star. The raw spectra are composed
of 248 measurements of the stellar
brightness between 0.85 and 5.0 $\mu$m 
with a typical resolution of 0.016 $\mu$m (in occultation mode, 
eight channels are used to encode timing data).
However, to save on data volume, these
data are usually co-added prior to transmission to earth,  producing 
``summed'' spectra consisting
of 31 spectral measurements  with a typical resolution
of 0.13$\mu$m. The raw data used in this analysis are 
the uncalibrated Data Numbers (DN) returned by the 
instrument. While these DN are linear measures
of the photon flux \citep{Brown04}, no attempt is made to convert these
data to absolute fluxes here, although a mean 
instrumental thermal background spectrum
has been subtracted from all the spectra
for each occultation.  A precise time stamp is appended to every 
spectrum to facilitate reconstruction of the occultation geometry.

Each occultation is geometrically navigated based on the positions
of the star (obtained from the Hipparcos catalog, and adjusted
to account for proper motion and parallax at Saturn) and the 
position of the spacecraft derived from the appropriate SPICE kernels.
This information was used to predict the apparent position (radius and inertial
longitude) of the star in Saturn's ring plane as a function of time 
in a planetocentric reference frame, taking into account stellar aberration. 
In nearly all cases, this estimate of the occultation geometry was
confirmed to be accurate to within a few kilometers using the known
radii of nearly circular gap edges in the outer A Ring from \citep{French93}. The exceptions
were the low-inclination stars $o$ Ceti and $\delta$ Virginis, for which
features could be tens of kilometers away from their nominal positions.
In these cases, the fiducial position of Saturn's pole was adjusted slightly 
(by at most 0.015$^\circ$) to bring these cuts into alignment with the other
occultations. (Note that such corrections were not
possible for the Rev 12 $o$ Ceti occultation, which only covered the F ring.) 

By the end of 2009, VIMS was able to detect the F ring, the
three dusty ringlets in the Encke Gap and the so-called ``Charming ringlet'' 
in the outer Cassini Division\footnote{This dusty ringlet is located at 119,940 km from Saturn
center, within the Laplace gap in the outer part of the Cassini Division.} 
in multiple occultations. The F ring
has been observed clearly most often because of its greater
optical depth, being detectable in all 87 occultation cuts where
the star passed behind that ring. The Encke Gap ringlets and 
Charming ringlet, by contrast, can only be clearly detected in cases
where the signal to noise is sufficiently high and when the
optical depth of the ringlet is sufficiently large because
the star either happened to pass behind a clump in the Encke Gap
ringlets or was observed at a very low ring opening angle.
In this analysis, we will only consider occultations where
the peak optical depth of these ringlets is at least five
times the standard deviation of the apparent optical depth variations 
outside the rings. This includes 24 cuts through the Encke Gap ringlets and
16 cuts through the Charming ringlet.

Tables~\ref{obstab}-\ref{charmtab} provide
lists of all the revelant occultation cuts, along with the occultation times, 
elevation angles and inertial longitudes of the observations and
the Data Numbers of the raw stellar signal. These
tables also indicate whether the spectra were returned
from the spacecraft in a summed or unsummed (full-resolution)
state, and provide various measures of the
rings' opacity, which are derived using the following procedures:
The data are first normalized so that the average signal levels
are unity at each wavelength in empty regions adjacent to each ring feature.
(138,000-139,000 km and  141,000-142,000 km from Saturn's center for the F ring,
133,510-133,540 km and 133,650-133,700 km the Encke Gap ringlets and
119,980-120,020 km for the Charming Ringlet). The resulting transmission
measurements $T$  are then transformed into line-of-sight slant optical depths
$\tau$ using the standard formula:
\begin{equation}
\tau=-\ln T.
\label{tau}
\end{equation}
Throughout this paper we will always use the symbol $\tau$ to designate
observed {\em slant} optical depths, never {\em normal} optical depths.
This is because the standard computed normal optical depths may not be valid 
for the numerous compact structures found in the F ring \citep{Esposito08, Murray08, Beurle10} and  the Encke Gap ringlets. Only for the Charming Ringlet, which  is relatively broad and shows little variation with longitude in images \citep{Hedman10C}, is the normal optical depth a sensible quantity. 
The tables therefore provide estimates of the 
maximum  {\it normal} optical depth for the cuts through the Charming Ringlet, and the 
maximum {\it slant} optical depths through the F ring and the Encke Gap ringlets
(the F-ring tables also provide  the minimum observed
transmission $T$).

\section{The Christiansen Effect in Saturn's dusty rings}

\begin{figure}
\resizebox{6in}{!}{\includegraphics{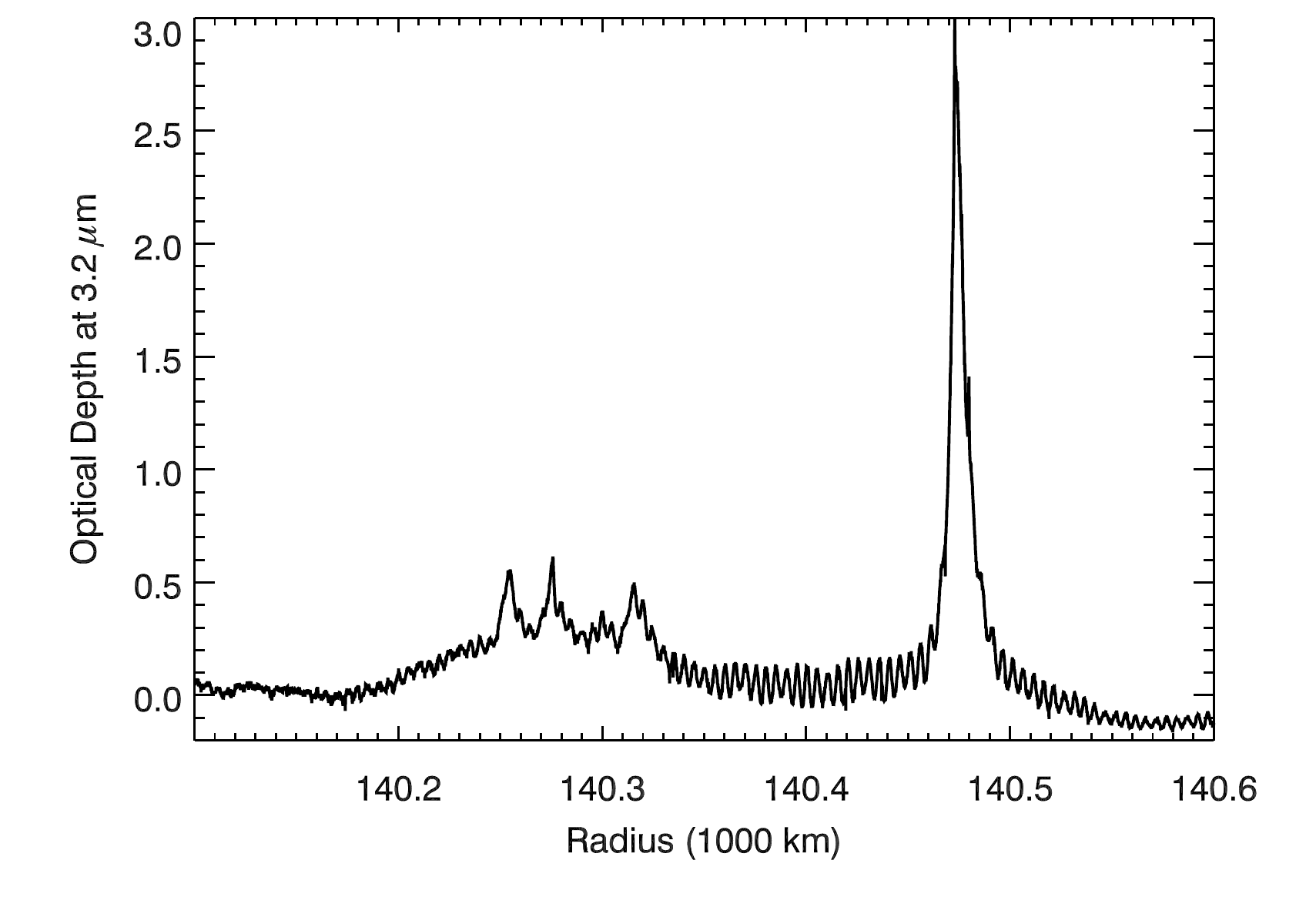}}
\caption{Optical depth profile of the F ring at 3.18$\mu$m as observed
in the Rev 46 $\alpha$Ori occultation. The rapid periodic variations
in optical depth are an instrumental artifact (see text).
Nevertheless, this is the highest signal-to-noise occultation
obtained at full spectral resolution. We derive the transmission
spectrum of the ring in Figure~\ref{specplot} by integrating over the strong peak at 140,470 km.}
\label{alporiprof}
\end{figure}

\begin{figure}
\resizebox{6in}{!}{\includegraphics{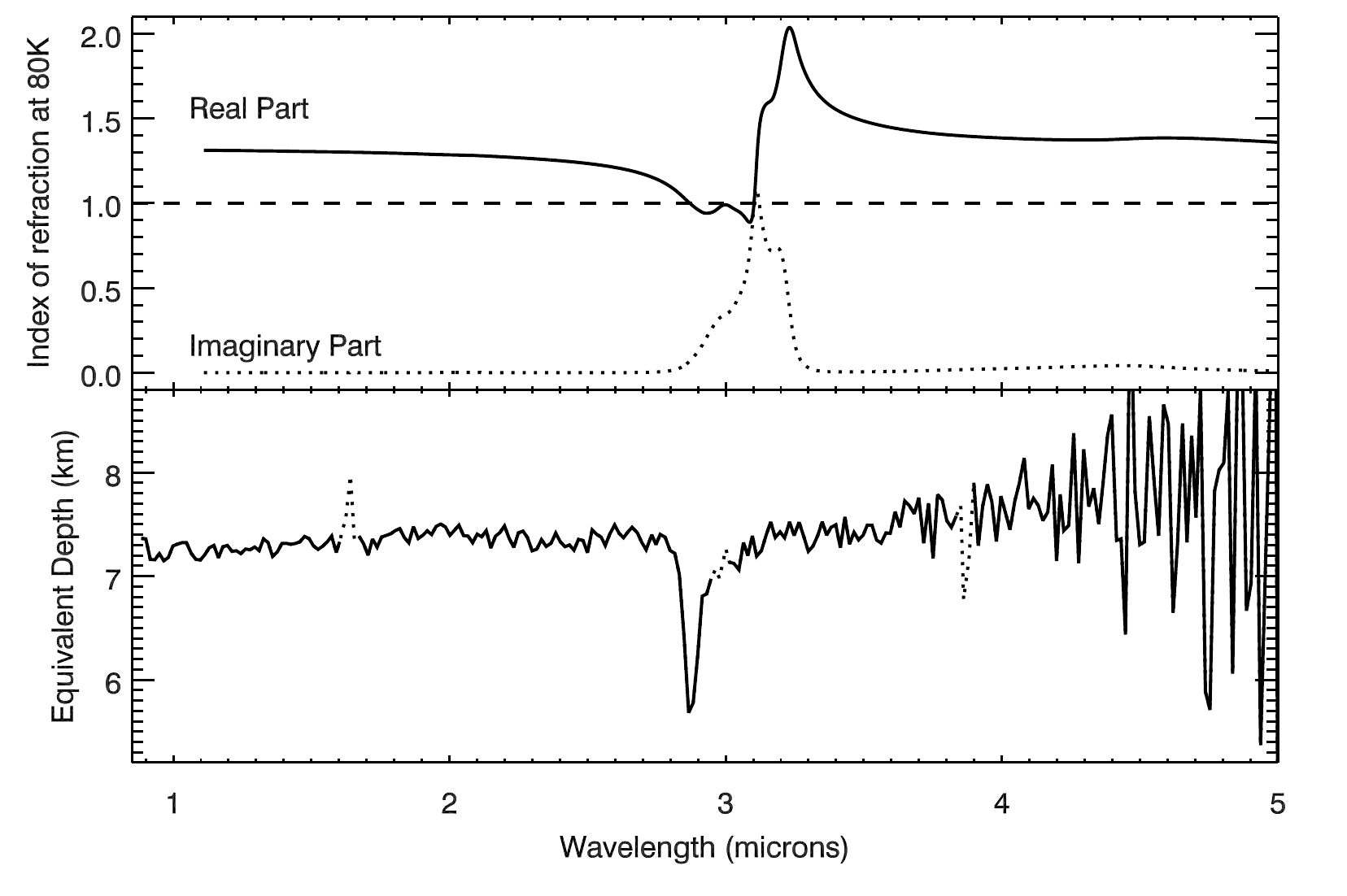}}
\caption{Wavelength-dependent opacity of the F ring core, as observed
in the Rev 46 $\alpha$Ori occultation. The top panel shows
the real and imaginary components of the index of refraction of crystalline water ice
at 80 K \citep{Mastrapa09},
while the lower panel shows the equivalent depth of the core of
the F ring as a function of wavelength. The dip in opacity at 2.87 $\mu$m
occurs where the real part of the index of refraction first approaches unity.
Dotted sections of the spectrum indicate boundaries between blocking filter segments, 
where the transmission estimates are less reliable.}
\label{specplot}
\end{figure}

The occultation data set that best illustrates the unique spectral
characteristics of these dusty ringlets  is the Rev 46 occultation
of $\alpha$ Orionis by the F ring. Of the handful of occultations done at full spectral resolution,
this one has the largest stellar signal and the lowest minimum transmission
(see Table~\ref{obstab}). Furthermore, the observed part of the F ring was 
within Saturn's shadow, so there is no contamination from scattered ringshine. 
The one problem with this occultation is that the observed stellar signal 
was not completely stable, but instead drifted and oscillated
by several percent even when there was no variation in the amount of
ring material occulting the
star (see Figure~\ref{alporiprof}), These variations probably occur because
the star happened to fall near the edge of the instrument's instantaneous 
field of view,
and thus the observed signal was especially sensitive to small jitters
in the spacecraft pointing. Even so, this data set  provides the 
highest signal-to-noise full-spectral-resolution measurements of the 
F-ring's transmission spectra to date.

In order to obtain a profile of ring opacity versus wavelength from these data, we convert the F-ring's optical depth profile at each wavelength into a 
single measure of opacity known as the normal equivalent depth \citep{French91}:
\begin{equation}
\mathcal{D}=|\sin B_{*}|\int \tau dr,
\label{eqdepth}
\end{equation}
where $B_{*}$ is the ring opening angle to the star. This radially-integrated 
quantity has units of length, and the factor of $|\sin B_{*}|$ may be
regarded as converting either the line-of-sight optical depth to normal optical depth for a flattened ring, or the radial integral into an integral in the sky plane for a spatially-diffuse ring. 
In principle, we could do the integral over the entire radial width of the F ring. 
However, in this case, we elected to integrate over only
the most optically-thick strand in the occultation 
(i.e. between radii of 140,400 and 140,550 km, 
see Figure~\ref{alporiprof}) in order to maximize the signal to noise.

Figure~\ref{specplot} shows the resulting plot of equivalent depth versus wavelength, along with the complex index of refraction for (crystalline) water ice derived by \citet{Mastrapa09}.
Naively, one might expect a peak in the ring's opacity around the peak
of the strong absorption band at 3.1 $\mu$m, but instead the most prominent feature in the data is a sharp dip in the opacity centered at 2.87 $\mu$m, just shortward of the 3.1 $\mu$m  band. This rather counter-intuitive minimum in opacity (or maximum in transmission) is almost certainly due to the Christiansen Effect \citep{Chris84, Chris85}, an optical phenomenon
that is observed in systems composed of many small particles like 
powders \citep{Prost68,EvZ06} or  ice clouds \citep{Arnott95, Liou98}. At most
wavelengths, the opacity of such materials is due to a combination
of absorption within, scattering from, and diffraction around 
the individual particles. However, near appropriately strong molecular
absorption bands the particles' optical properties become
strongly wavelength-dependent, and there can even be wavelengths where
the particles' index of refraction has a real part $n_r$ close to 
that of the background medium while the imaginary part  $n_i$ is still well less
than unity.
At these so-called Christiansen wavelengths, surface scattering and diffraction are
strongly suppressed, and provided the particles are sufficiently small, this produces 
a dip in the overall opacity \citep{Hapke93, EvZ06, Vahidinia11}.

Since the primary constituent of Saturn's rings is crystalline water ice \citep{Cuzzi09},
and the ring particles are dispersed in free space, the F-ring's Christiansen wavelengths 
should occur wherever crystalline water ice has both $n_r\simeq1$ and $n_i<<1$. 
Examining the optical constants for crystalline water ice 
plotted in Figure~\ref{specplot}, we find the indices of refraction vary dramatically in the vicinity of the strong 3.1 $\mu$m absorption band,
with $n_r$ falling below unity between 2.87 and 3.10 $\mu$m.
In this range of wavelengths, ice only has $n_r\simeq1$ and $n_i<<1$
at 2.87 $\mu$m. Thus the deepest part of the observed opacity dip falls very close to the Christiansen wavelength for crystalline water ice
\footnote{Amorphous water ice at 80 K has a Christiansen wavelength
at 2.84 $\mu$m, which is inconsistent with the observed feature \citep{Vahidinia11}. This observation thus provides further evidence for the crystallinity of the ice in the rings \citep{Cuzzi09}.}, providing
strong evidence that the Christiansen Effect is indeed responsible for this feature.

\begin{figure}
\resizebox{6in}{!}{\includegraphics{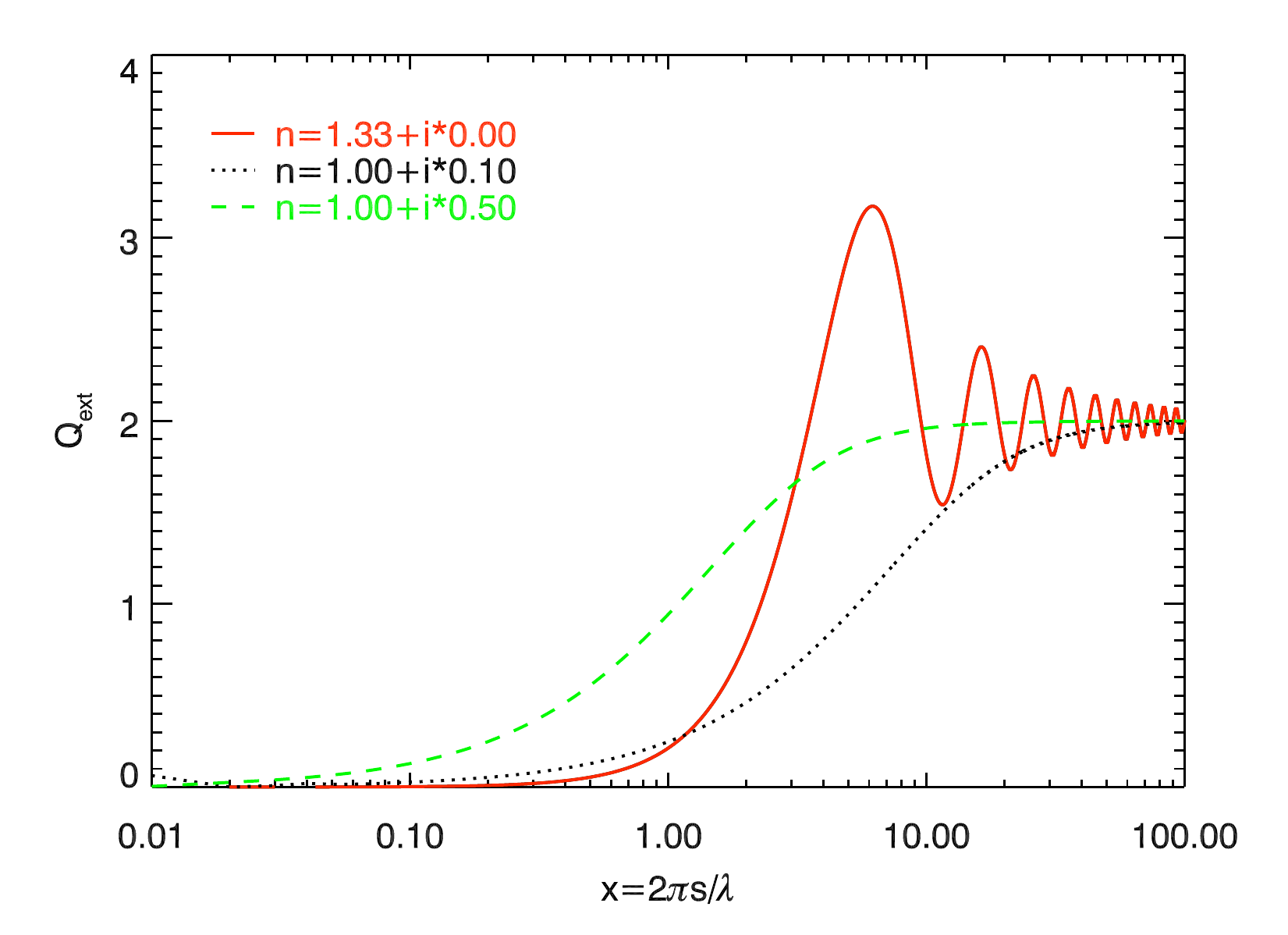}}
\caption{Plot  of the extinction coefficient $Q_{ext}$ of spherical
grains with different indices of refraction as a function of $x=2\pi s/\lambda$. 
The $n=1.33+i*0.0$ case corresponds to ice outside of strong molecular absorption bands,
while the other two cases reflect the variations in the optical
constants within the strong 3.1 $\mu$m absorption band.}
\label{qext}
\end{figure}

The opacity dip at the Christiansen wavelength is only observed
in the F ring and other similarly ``dusty'' ring features (see below). 
It is not seen in the icy main rings because the particles
in these rings are all millimeters in size or larger \citep{Cuzzi09}, and
the Christiansen Effect is only important for wavelength-sized particles. 
This can be most easily understood by
considering the extinction coefficient $Q_{ext}$, which is
defined as the ratio of a particle's optical cross section 
(including loses due to both scattering and absorption) to 
its physical cross section. In general, the relationship between
this parameter and the measured opacity of a ring 
is rather complex and depends on the width of the ring particles' 
forward scattering lobe, as well as the apparent size of the 
observed ring feature and the instrument's field of view  
\citep{Cuzzi85, FN00}. However, for rings 
composed primarily of sub-millimeter particles (which have broad 
forward-scattering diffraction lobes) the measured opacity 
is simply proportional to the appropriately weighted average of 
$Q_{ext}$ over all the particles in the ring.
For dielectric spheres $Q_{ext}$ can be computed analytically \citep{vandeHulst}:
\begin{equation}
Q_{ext}=2-4e^{-\chi\tan\beta}\left(\frac{\cos\beta}{\chi}\right)\left[\sin(\chi-\beta)
+\frac{\cos\beta}{\chi}\cos(\chi-2\beta)\right]+4\frac{\cos^2\beta}{\chi^2}\cos(2\beta).
\end{equation}
Here $\tan\beta=n_i/(n_r-1)$ and $\chi=(n_r-1)4\pi s/\lambda$,
$n_r$ and $n_i$ being the real and imaginary parts
of the particle's refractive index, while $s$ is the particle radius 
and $\lambda$ is the wavelength of the incident light. For very small particles
($s<<\lambda$), $Q_{ext}$ approaches zero, while for large 
particles ($s>>\lambda$), $Q_{ext}$ asymptotes to 2 regardless
of the assumed optical constants. However, at intermediate
particle sizes where $x=2\pi s/\lambda \sim 1$, $Q_{ext}$ 
can vary dramatically depending on the assumed indices of refraction 
(see Figure~\ref{qext}). Since the opacity of a diffuse cloud of particles is 
proportional to the appropriately weighted average of $Q_{ext}$, 
the precise values of the  particles' optical constants will only be relevant
to a ring's optical depth if there are sufficient particles of size $s \sim \lambda$.
Thus the opacity of Saturn's main rings, where $s>>\lambda$ for all particles,
should be almost independent of wavelength, as observed, and only
dusty rings should exihibit spectral features in their transmission spectra.

On a more detailed level, the magnitude of the opacity variations
in the vicinity of the Christiansen wavelength can help constrain the
particle size distributions in these dusty rings. Indeed, the curves
in Figure~\ref{qext} provide  insights into which aspects of the particle 
size distribution the opacity measurements probe.
First consider the curve corresponding to $n=1.33+i*0.0$, 
the typical index of refraction of water ice outside the strong absorption
bands. In this case the primary sources of opacity are surface scattering
and diffraction, and the periodic variations in $Q_{ext}$ seen
in this curve arise from interference among light rays taking different paths 
around or within the grains. These wiggles are very sensitive to
the exact size and shape of the particles and are therefore likely to be
washed out for any collection of realistically-shaped grains. However, 
even if ignore these ripples, there are still clear systematic differences
between this curve and the two others, which are more representative
of the optical constants near the strong 3.1 $\mu$m absorption band.

The $n=1.00+i*0.1$ curve corresponds to water ice near
the Christiansen wavelength of  2.87 $\mu$m (see Figure~\ref{specplot}). 
The extinction in this case is largely due to bulk absorption and is significantly lower 
than for the $n=1.33+i*0.0$ case in the range 
$1<x=2\pi s/\lambda<10$. Thus, ice-rich particles between 1 and 10
$\mu$m in diameter should have a {\em reduced} opacity at 2.87 $\mu$m
compared to that at wavelengths far from the absorption band. 
We can therefore interpret the observed opacity dip in the F ring 
at 2.87 $\mu$m as evidence for a significant
population of  1-10 $\mu$m-sized particles in this ring.
 
On the other hand, the $n=1.00+i*0.5$ curve corresponds
to the $Q_{ext}$ of water ice around 3.1 $\mu$m, near the center of the
absorption band. In this situation, the predicted $Q_{ext}$ is noticably higher 
than the $n=1.33+i*0.0$ case when  $x=2\pi s/\lambda<2$. Thus 
particles smaller than 2 $\mu$m in diameter would tend to have a 
{\em higher} opacity at 3.1 $\mu$m than they do at other wavelengths. 
This would produce a peak in
the opacity spectrum, which is {\em not} observed, and therefore
suggests that sub-micron grains do not dominate the opacity of the F ring.
This is consistent with the lack of a pronounced slope in the
transmission spectrum outside of
the  dip at 2.87 $\mu$m. If sub-micron grains were common in
the ring, then far from the strong absorption band we would
expect $Q_{ext} \propto (s/\lambda)^2,$ which would produce
a steep slope in the transmission spectrum which is not seen
either here or in earlier ground-based occultation data \citep{Bosh02}. 

\begin{figure}
\resizebox{6in}{!}{\includegraphics{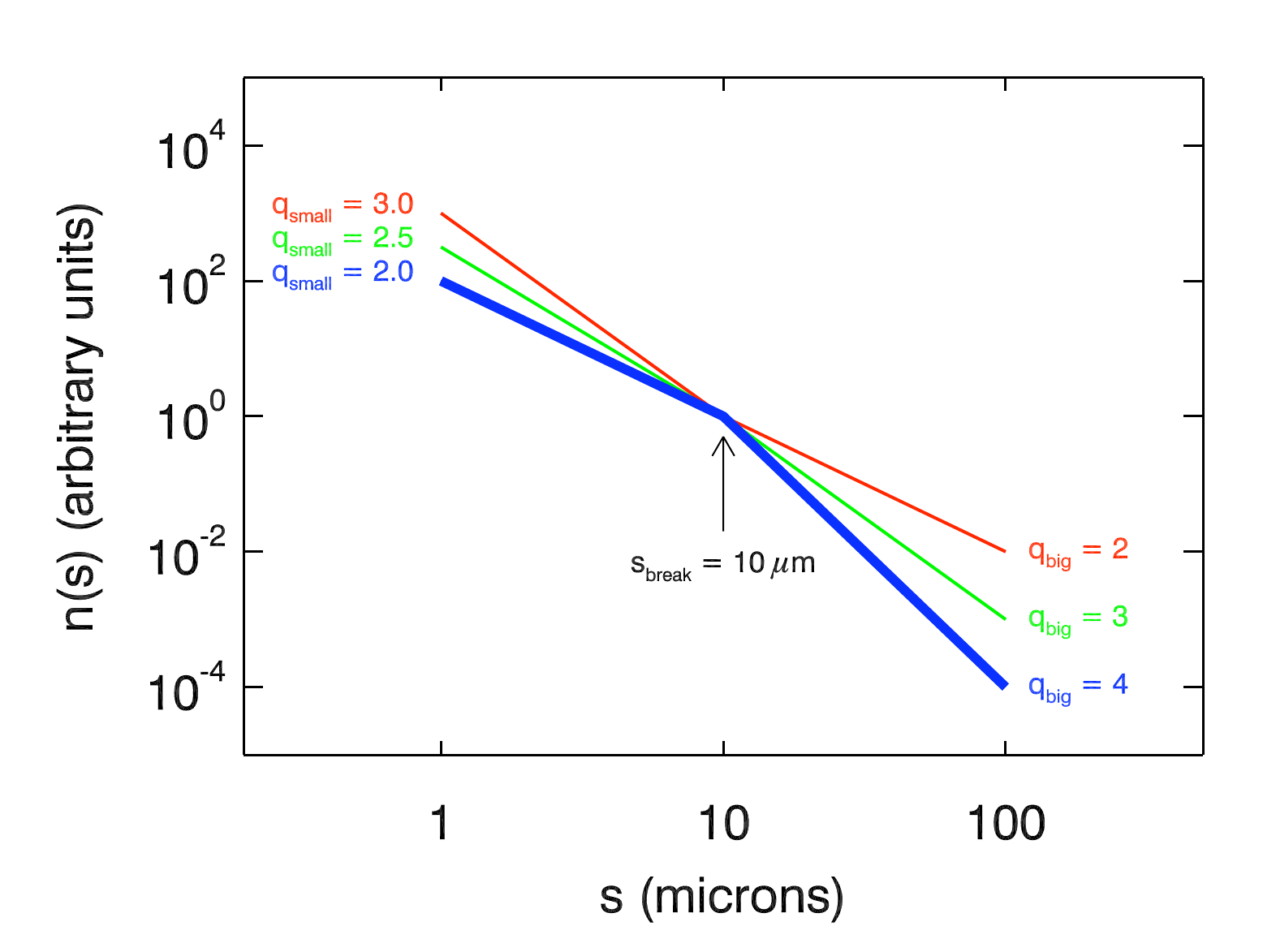}}
\caption{Diagram illustrating the parameters used to describe
broken power-law size distributions, which include two slopes
$q_{small}$ and $q_{big}$, as well as a break point $s_{break}$.}
\label{pldiag}
\end{figure}

\begin{figure}
\resizebox{6in}{!}{\includegraphics{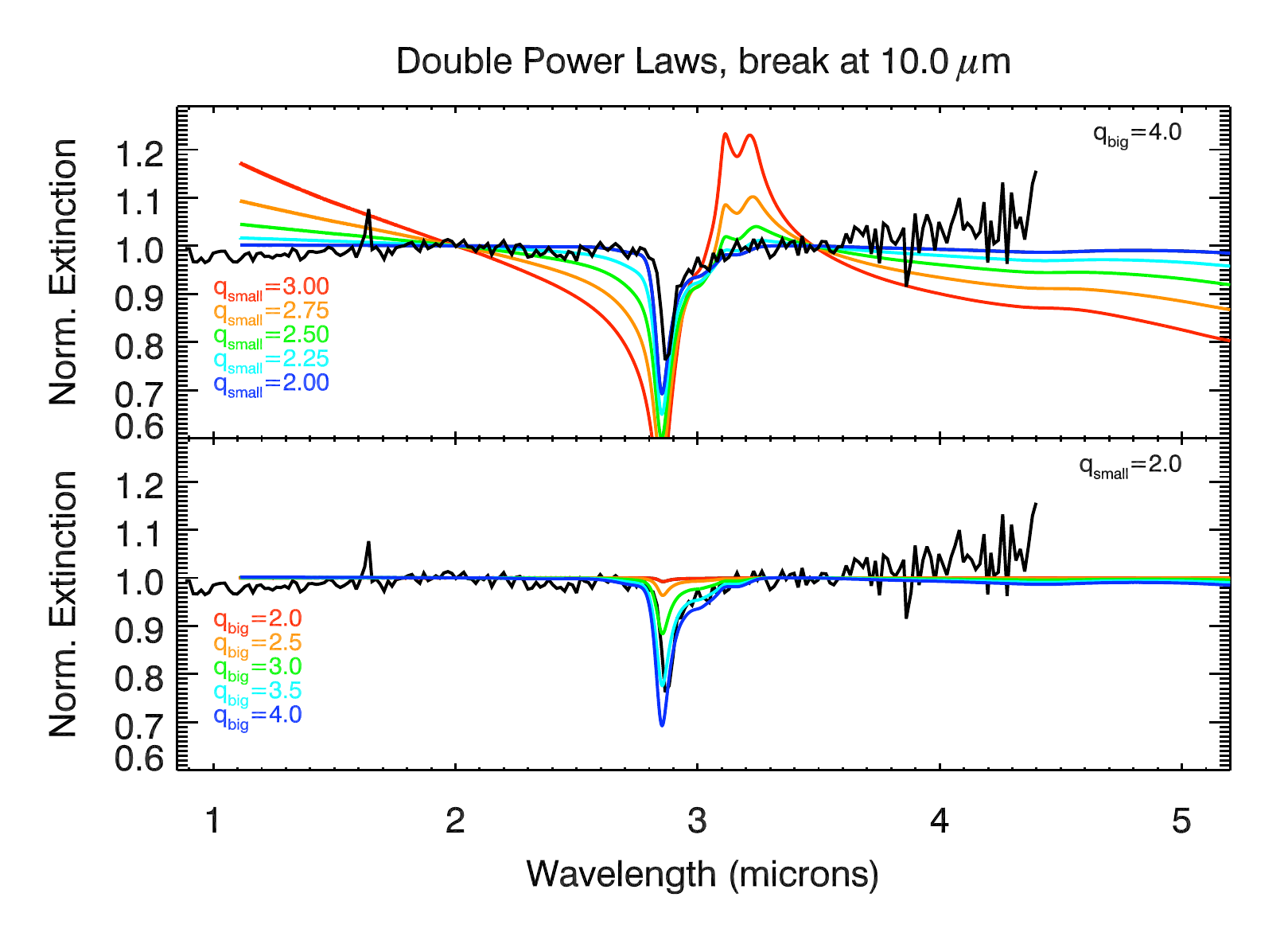}}
\caption{Model extinction spectra for different particle size distributions
compared with the observed spectra derived from the  Rev 46 $\alpha$ Orionis
occultation. All spectra have been scaled to unity at 2 $\mu$m in order
to facilitate comparisons. All model spectra are for broken power laws with
the break at  a particle radius of 10 $\mu$m and a maximum particle
size of 1 mm. All grains are assumed to be composed of pure crystalline water ice with
optical constants given in \citet{Mastrapa09}.
In the top panel, all the model distributions have a power-law index of 4 above the
break, but a range of indices between 2 and 3 below the break. Note
that as the number of small particles increases, the spectrum outside the
2.87-$\mu$m opacity dip develops prominent slopes that are not observed
in the data. In the lower panel, all the model size distributions
have indices of 2 below the break, but a range of indices above
the break. In this case, we see that if there are too many large particles,
then the opacity dip will become diluted and is too small compared
to the observations.}
\label{plmod}
\end{figure}

The above considerations suggest that only
a rather limited set of particle size distributions
will be able to reproduce the observed
transmission spectrum. To test this 
supposition, we computed
the predicted transmission
spectra for various populations of 
particles using a Mie scattering code, assuming
that all the particles are composed of pure crystalline
water ice at 80 K. Initial investigations indicated that
neither simple power-laws  nor narrow Hansen-Hovenier 
distributions could reproduce the observed
transmission spectra. We therefore considered
slightly more complex size distributions, and found
that certain broken power laws could reproduce
many of the salient features of 
the observed transmission spectra.
In these models the particle size distribution 
follows a power law 
$n(s) \propto s^{-q_{small}}$ up to a critical size
$s_{break}$, above which the size distribution changes to 
a different (steeper) power law $n(s) \propto s^{-q_{big}}$ (see Figure~\ref{pldiag}).  
Figure~\ref{plmod} shows the calculated extinction spectra
for a range of models with different  values of
$q_{big}$ and $q_{small}$, but the same 
value of $s_{break}=10 \mu$m and an assumed maximum particle
size of 1 mm. As expected, increasing  $q_{small}$ ---which increases the
fraction of  micron-sized and smaller particles in the ring---  causes the extinction
spectrum to develop a prominent  slope at short wavelengths
and a peak at 3.1$\mu$m. Since these features
are not seen in the F-ring data, $q_{small}$ needs to 
be rather low (around 2 assuming $s_{break}\sim 10 \mu$m).
On the other hand, decreasing $q_{big}$ ---which increases the fraction
of very large particles in the ring--- tends to dilute the 2.87$\mu$m opacity dip.
Thus to produce an opacity dip of the appropriate magnitude
$q_{big}$ must be fairly high (around 3.5 assuming $s_{break}\sim 10 \mu$m).
Together, these findings indicate that the size distribution 
must have a rather sharp break (with $q_{small} \sim 2$ and $q_{big} \sim 3.5$) to 
reproduce the observed transmission spectra. Such
a break is consistent with other spectral data
for the F ring \citep{Vahidinia11}. 

Of course, such transmission spectra alone cannot uniquely 
determine the particle size distributions of these rings. For example,
different assumed values of $s_{break}$ lead to somewhat
different preferred values of $q_{big}$. Indeed,  
these data can only place limits on the fraction of the ring particles
larger than 10 $\mu$m across, and do not strongly constrain 
the typical size of these larger particles. Furthermore, the ring particles
are not spheres of pure water ice, and the detailed
microstructure of the grains can alter  
the depth and location of the dip and complicate efforts
to quantitatively constrain the particle size distribution \citep{Vahidinia11}. 
Indeed, the observed dip occurs at slightly longer wavelengths 
than predicted by the simple models used here. Nevertheless, these 
transmission spectra still provide a unique resource for exploring 
{\em variations} in the particle size distributions of these dusty rings.

\section{Comparisons among different dusty rings}

The F ring and Encke Gap ringlets exhibit abundant structure and 
large variations in optical depth, so it is natural to ask whether
their particle size distributions vary as well.
By virtue of their high spatial resolution, the occultation data provide
a unique avenue for investigating and quantifying any differences
in the particle size distributions within or among these dusty rings. 
However, in practice we can only obtain a limited amount of spectral
information from most of the occultations. In many 
occultations the total light observed by VIMS consists of a 
combination of both the desired starlight and sunlight reflected 
by the rings.  This background ringshine is often sufficiently 
bright outside the strong 3.1 $\mu$m band that we cannot reliably 
measure any slope in the continuum transmission spectrum.
On the other hand, the rings are always very dark at wavelengths
near  3.1 $\mu$m, so the size and shape of the opacity
dip are uncontaminated by ringshine. Unfortunately, most of the 
occultations were spectrally summed, so the spectral 
resolution of the measurements is insufficient to 
discern any changes in the width or shape of the opacity dip, 
leaving the magnitude of the dip as the only spectral
feature that can be reliably determined for most of the occultations.
Fortunately, even this single spectral parameter is sufficient to
provide useful information about the variability of the particle
size distribution in these dusty rings.

\begin{figure}
\resizebox{6in}{!}{\includegraphics{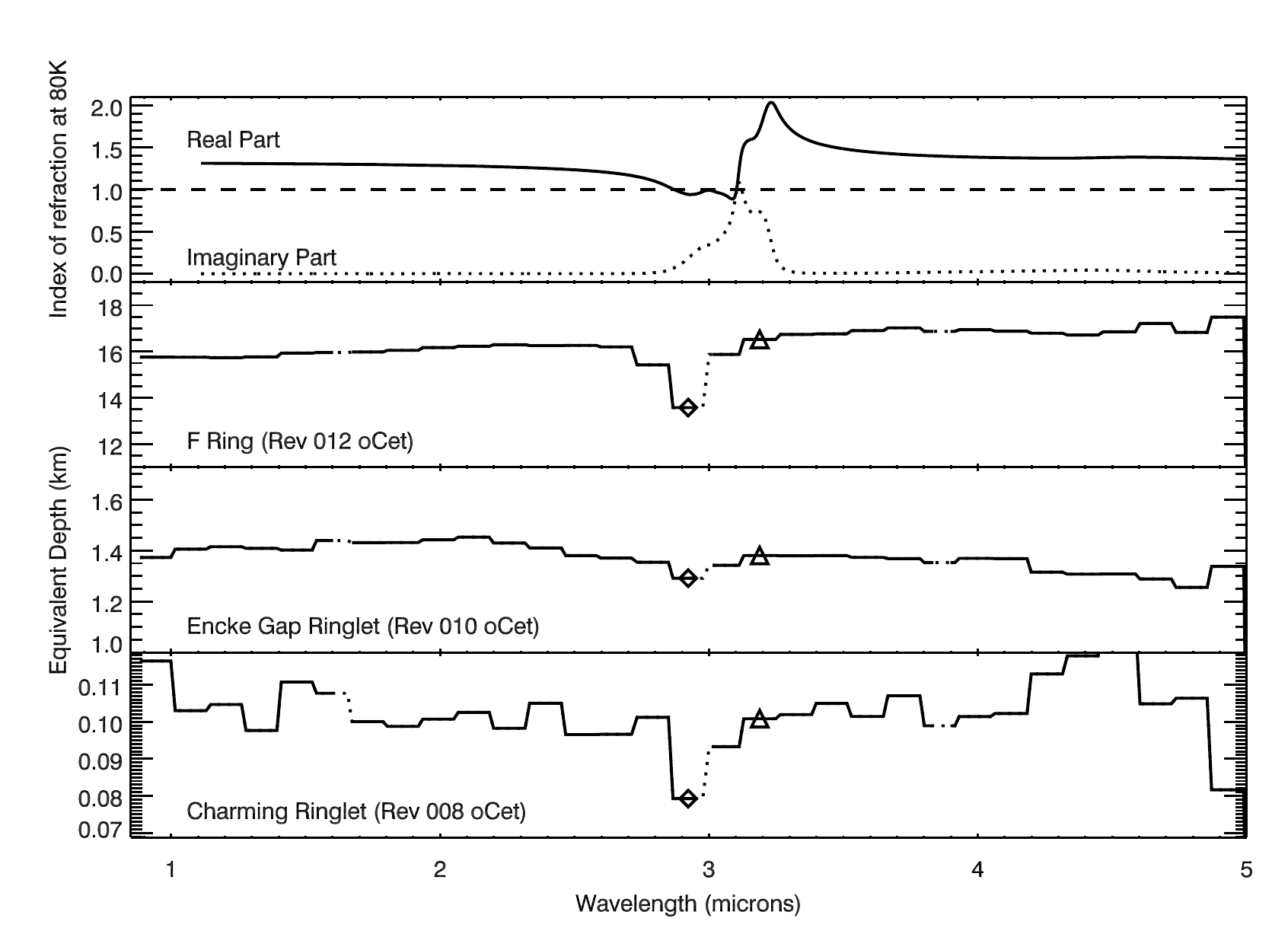}}
\caption{The opacity of the F-ring, an Encke Gap ringlet and the 
Charming ringlet measured during occultations of the star $o$ Ceti.
The data here are all spectrally summed and therefore have lower 
resolution than the $\alpha$Ori data shown in Figure~\ref{specplot}.
The diamond and the triangle mark the channels used to estimate
the strength of the opacity dip (see text). All three spectra are of parts of
the rings inside Saturn's shadow, and are computed by integrating
over the radial ranges discussed in the text. Dotted lines in the spectra
indicate filter gaps, as in Figure~\ref{specplot}. The indices of refraction
are for crystalline water ice at 80 K \citep{Mastrapa09}.}
\label{ocetspec}
\end{figure}

Figure~\ref{ocetspec} illustrates how the opacity dip
appears in the spectrally summed data. While the
shape of the dip is no longer resolved, the spectral channel
covering the range 2.87-2.98 $\mu$m (marked by a diamond)
shows an equivalent depth that is clearly below its neighbors. By contrast,
the spectral channel covering 3.13-3.25 $\mu$m
(marked by a triangle) is clearly outside this dip. 
The main rings are observed to be quite dark at both these wavelengths
due to absorption by water ice,
so we can reliably quantify the magnitude
of the opacity dip using the measured optical depths in 
these two summed spectral channels.
Given that that the detailed microstructure of the dust grains
can affect the depth and morphology of the opacity dip \citep{Vahidinia11},
we will avoid any attempt to convert the absolute
value of the opacity dip in the summed data into a constraint on the particle
size distribution in the ring.\footnote{This caution is also justified 
because the last two spectral channels included in the summed 
2.87-2.98 micron channel fall within a filter gap on the focal plane.
While no evidence has been found that this gap allows light from
other wavelengths to enter these channels, the absolute 
photometry could be compromised  (see also Vahidinia {\em et al.} 2011).} Instead, we will concentrate on trends in the depth of the opacity dip 
among and within the different rings.


For the rest of this analysis, we will quantify 
the magnitude of the opacity dip observed in the various
dusty rings using the optical depth ratio: 
\begin{equation}
\rho=\frac{\tau_{2.9}}{\tau_{3.2}},
\end{equation}
where $\tau_{2.9}$ and $\tau_{3.2}$ are the measured
slant optical depths in the summed spectral channels
covering the ranges 2.87-2.98 $\mu$m and 3.13-3.25 $\mu$m.
These optical depths are computed from the appropriately
normalized transmission profiles using the procedures 
described in Section~\ref{obssec} above, so for each occultation
cut through the rings we can derive $\rho$ as a function of radius. 

While $\rho$ itself is useful for studying variations in the opacity dip's 
strength within a single occultation cut, for comparing data from 
different cuts through different rings it is also worthwhile to have
a radially-averaged optical depth ratio for each
occultation cut through each ring feature. Simple radial averages
are not appropriate in this situation because the value of $\rho$ becomes ill-defined when the optical depth is low. Thus we instead compute a {\it weighted} 
average of  $\rho$, where the weight is simply the optical depth at 3.2 $\mu$m:
\begin{equation}
\langle\rho\rangle=\frac{\int \tau_{3.2} \rho dr}{\int \tau_{3.2} dr}
	=\frac{\int \tau_{2.9} dr}{\int \tau_{3.2} dr}=
	\frac{\mathcal{D}_{2.9}}{\mathcal{D}_{3.2}}.
\end{equation}
where $\mathcal{D}_{2.9}$ and $\mathcal{D}_{3.2}$ are the ring feature's  
equivalent depths at the two wavelengths (see Equation~\ref{eqdepth}). Thus
$\langle\rho\rangle$ is the ratio of the equivalent depths at 2.9$\mu$m and
3.2$\mu$m, which is the most sensible average statistic for a narrow ring.

The specific radial ranges used in the calculation of $\langle\rho\rangle$
differ for each ring feature. They are 139,000-141,000 km for the F ring, 133,450-133,510, 133,540-133,650 and 137,000-133,730 for the inner, 
central and outer Encke gap ringlets, and 119,880-119,980 km for the 
Charming Ringlet. The radial range for the F ring is deliberately broad in order
to include all of its multiple strands and encompass its substantial orbital
eccentricity \citep{Bosh02, Murray08}. When doing each integration we 
deliberately exclude all data where $\tau_{3.2}$ is less than 
5 times $\sigma_{\tau}$, the standard deviation of $\tau_{3.2}$ in the empty regions adjacent to the ring feature. We also exclude any data where the transmission falls below 0.1 (i.e. $\tau>2.3$), in order to avoid regions where the optical depth may be saturated.
The resulting values of $\langle\rho\rangle$ are recorded in Tables~\ref{obstab}--~\ref{charmtab}, but we will use both the localized, single-sample 
values of $\rho$ and the radially-averaged quantities $\langle\rho\rangle$
in the discussions below.

\begin{figure}
\resizebox{6in}{!}{\includegraphics{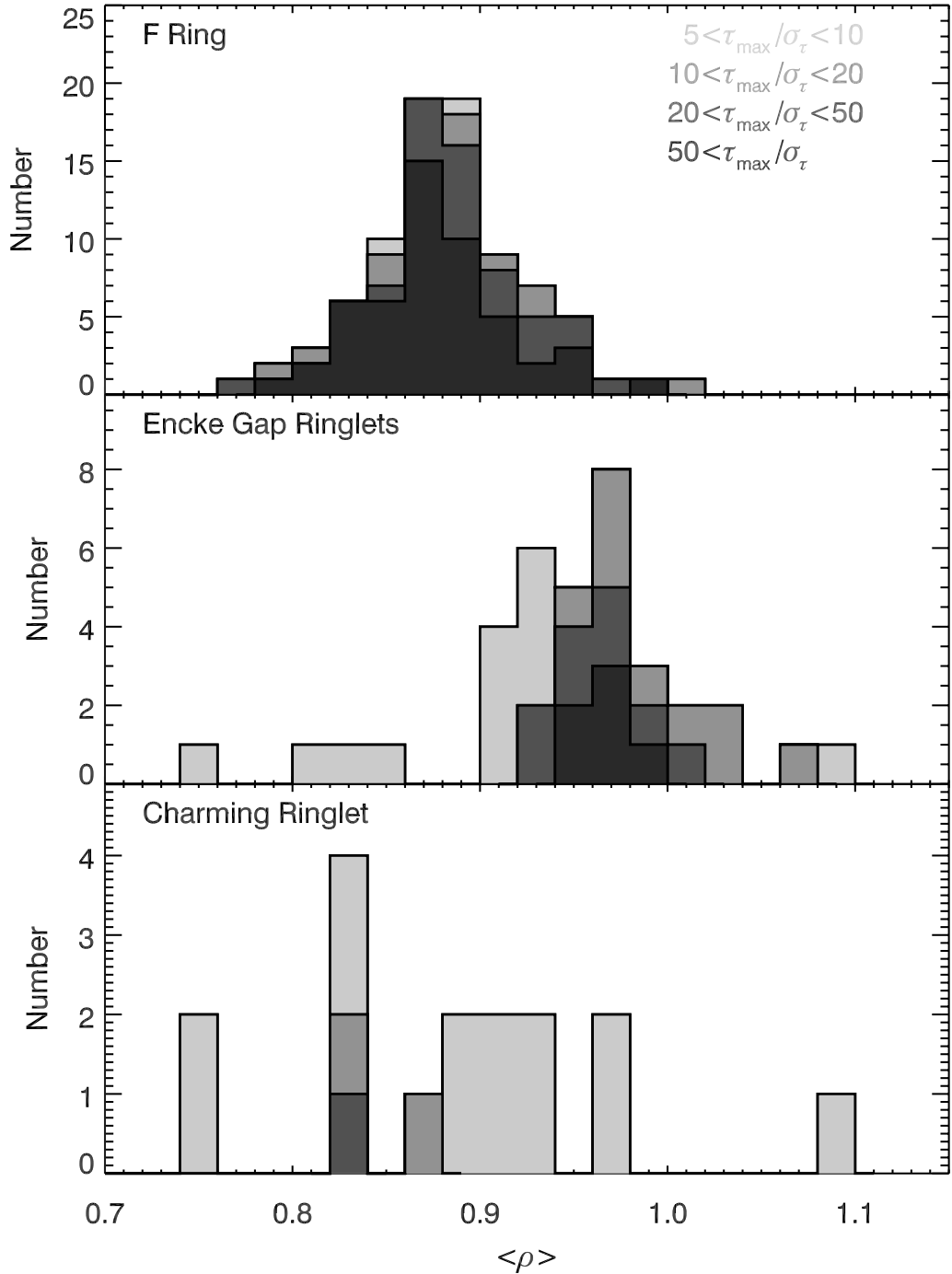}}
\caption{Histograms showing the distribution of radially-averaged
$\langle\rho\rangle$ values for the F ring, Encke gap Ringlets and
the Charming Ringlet.The different gray levels correspond
to values of the ratio $\tau_{max}/\sigma_\tau$ as indicated, with
lighter-toned histograms stacked on top of (not overlapping)
the heavier-toned histograms }
\label{rhodist}
\end{figure}

\begin{figure}
\resizebox{6in}{!}{\includegraphics{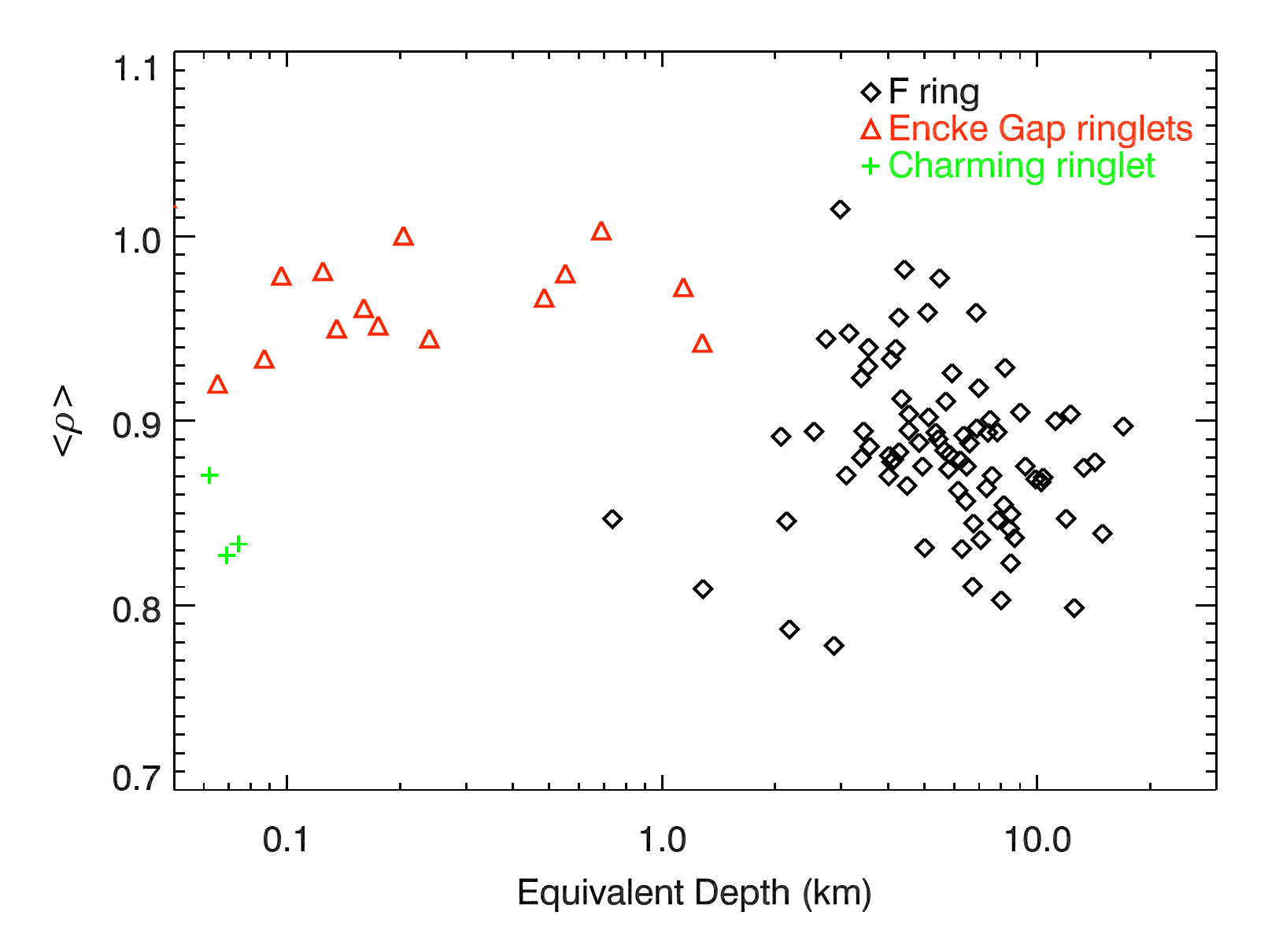}}
\caption{Scatter plot showing the  radially-averaged 
$\langle\rho\rangle$ values versus the equivalent depth of the ring at 3.2 $\mu$m.
Only data with $\tau_{max}/\sigma_\tau>10$ shown. The F ring
data are the black diamonds, the Encke Gap ringlet data are
the red triangles, and the Charming Ringlet data are
the three green plusses.}
\label{rhoscat}
\end{figure}

\begin{figure}
\resizebox{6in}{!}{\includegraphics{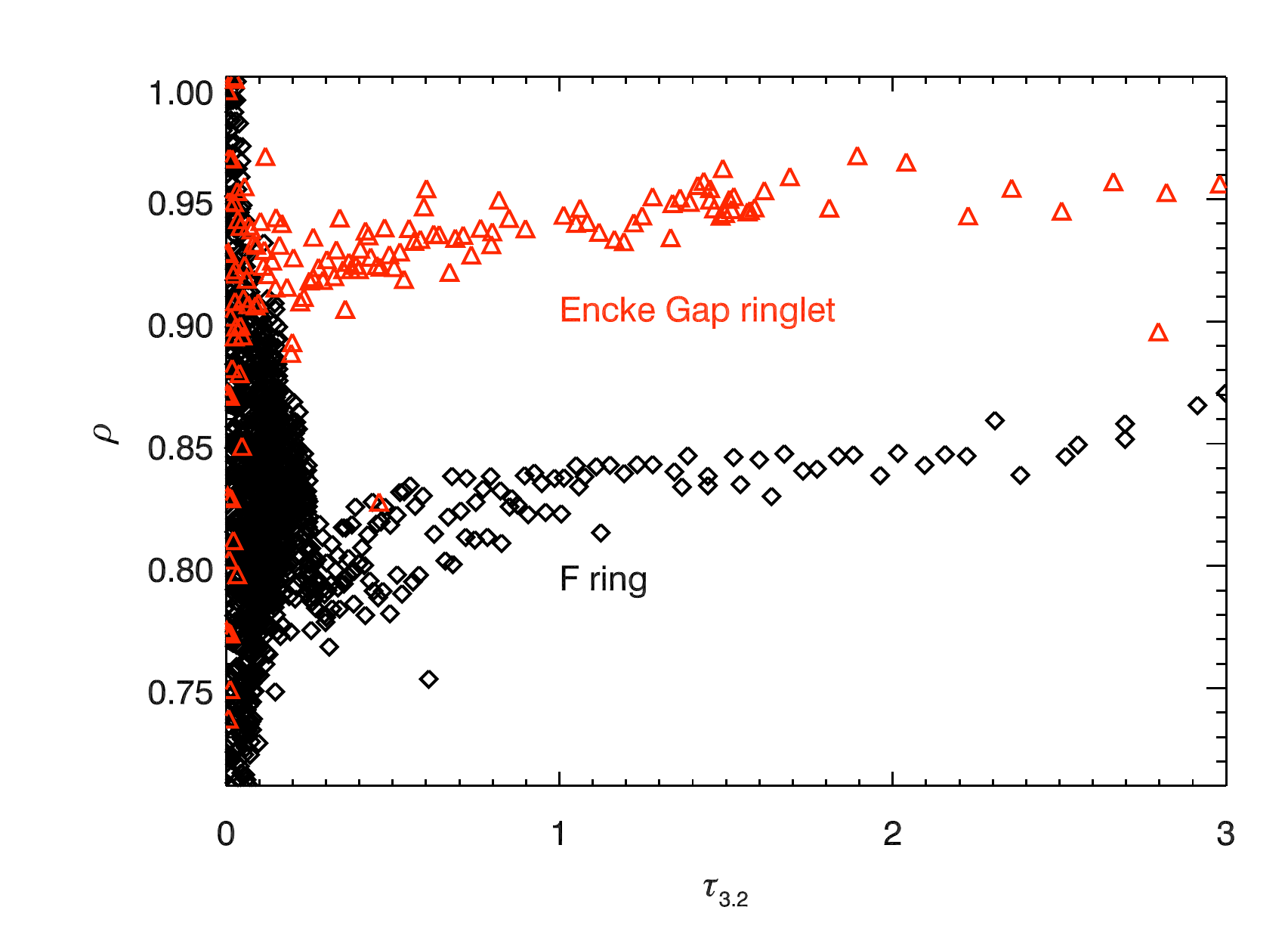}}
\caption{Scatter plot showing the single-sample values of $\rho$ versus slant optical
depth for the F ring and central Encke Gap ringlet derived from the
Rev 10 $o$ Ceti ingress occultation, which happened to cut
through a particularly dense clump in the Encke Gap ringlet. 
Note that at low optical depths, $\rho$ becomes ill-defined
leading to a large scatter in the measurements.
While both features show a slightly increasing trend in $\rho$ with
optical depth, the Encke Gap ringlet has a consistently higher $\rho$
than the F ring over a broad range of $\tau$.}
\label{rhotrend}
\end{figure}

Figure~\ref{rhodist} shows histograms of the derived 
$\langle\rho\rangle$ values for the various ring features. The shading in the
histograms represent data with different signal-to-noise ratios,
parametrized as the ratio of the peak optical
depth at 3.2 $\mu$m ($\tau_{max}$) to the standard deviation of the optical depth
values ($\sigma_\tau$) in nearby empty regions. Note that in both the Encke Gap and Charming
Ringlet distributions, the data with  $\tau_{max}/\sigma_\tau$ between 5 and 10 
are obviously more scattered than the data with $\tau_{max}/\sigma_\tau$ 
greater than 10. This implies that the signal-to-noise
is too low to obtain reliable $\langle\rho\rangle$ estimates when  $\tau_{max}/\sigma_\tau <10$.
We therefore tabulate in Tables~\ref{obstab}-~\ref{charmtab} the $\langle\rho\rangle$
values only for those observations where $\tau_{max}/\sigma_\tau >10$.

Focusing  exclusively on the higher significance data in Figure~\ref{rhodist},
it is apparent that the different dusty rings have
different distributions of $\langle\rho\rangle.$ 
The F-ring distribution peaks around 0.9, but extends from below
0.8 to about 1.0. The three clear measurements of $\langle\rho\rangle$ from
the Charming Ringlet all fall between 0.8 and 0.9, and therefore overlap
the F-ring distribution. However, the Encke Gap ringlets seem to have 
$\langle\rho\rangle$ values between 0.95 and 1, which is clearly higher
than typical values for the F ring.

Figure~\ref{rhoscat} shows the $\langle\rho\rangle$ values versus  the equivalent depth for the different rings (computed using Equation~\ref{eqdepth} and the same radial ranges as employed in the calculations of $\langle\rho\rangle$). This plot demonstrates that the spectral differences between the Encke Gap ringlets and the F ring cannot be 
entirely ascribed to
differences in these features' average optical depths. Even though the F ring's equivalent depth
is almost always higher than the Encke Gap ringlets, there are cases where the
equivalent depth of these features are comparable to each other, and 
even in these situations the $\langle\rho\rangle$ values of the Encke
Gap ringlets are systematically higher than those of the F ring.
The Encke Gap ringlets and the F ring would therefore appear 
to have systematically different particle size distributions (see also Figure~\ref{rhotrend} below). Since there is no evidence for a spike in the optical depth at 3.1 $\mu$m in any of the dusty rings 
(see Figure ~\ref{ocetspec}), they all are probably depleted 
in sub-micron particles (cf. Figure~\ref{plmod}). 
However, the higher $\langle\rho\rangle$ values in the Encke Gap ringlets
(corresponding to a weaker opacity dip at 2.87 $\mu$m) implies that
these ringlets  have a smaller fraction of particles in the 1-10 $\mu$m  range. 
The size distribution of grains larger than 10 $\mu$m in the Encke Gap 
ringlets is therefore probably not as steep as it is for the typical F ring or the
Charming Ringlet.

\begin{figure}
\centerline{\resizebox{5in}{!}{\includegraphics{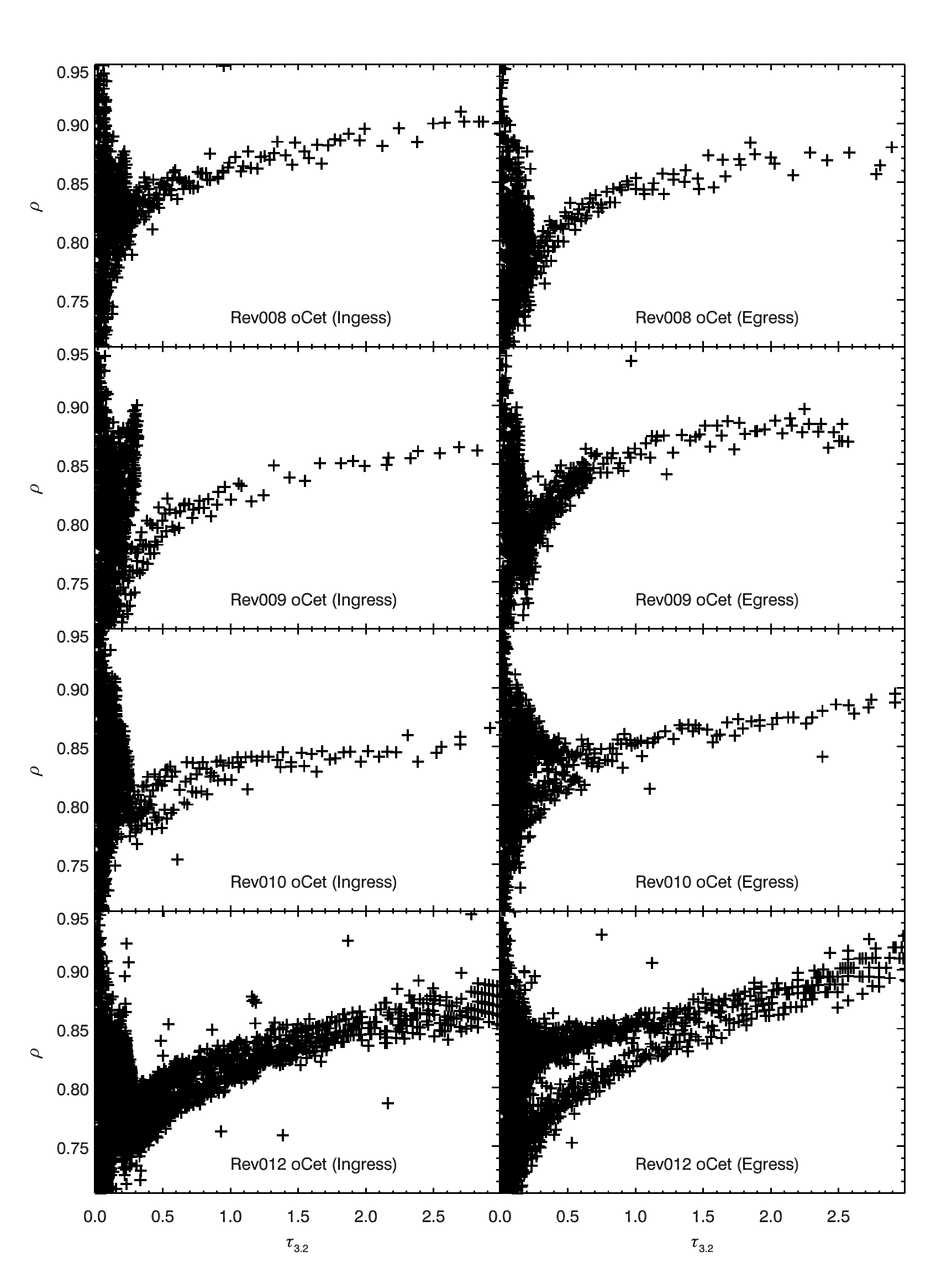}}}
\caption{Trends in the single-sample $\rho$ values versus slant optical depth for the eight $o$ Ceti occultations by the F ring.
Note that at low optical depths, $\rho$ becomes ill-defined
leading to a large scatter in the measurements.
All these profiles show the same basic trend of increasing $\rho$ with increasing 
optical depth, consistent with the idea that larger particles are concentrated near the
core of the ring. However, there are also significant variations in the shape of
this trend among the different occultation profiles (The Rev 12 occultation track reached a minimum radius interior to the F ring, resulting in an exceptionally large
number of samples within this ring).}
\label{ocettrend}
\end{figure}

\begin{figure}
\resizebox{6in}{!}{\includegraphics{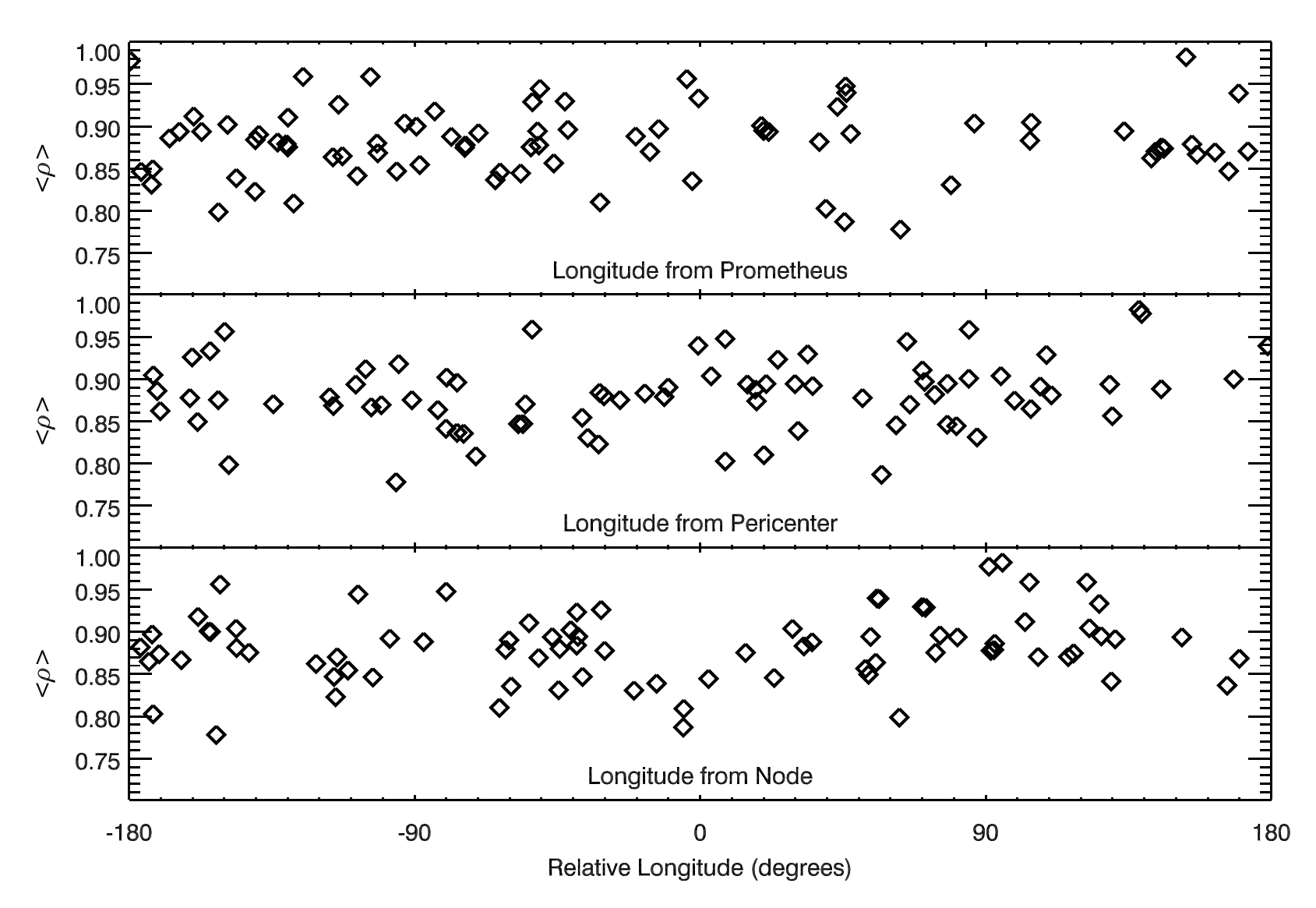}}
\caption{Scatter plots of the radially-averaged $\langle\rho\rangle$ values for the F ring
versus longitude relative to Prometheus and the  F-ring's pericenter and node
(Model 12 of Albers {\it et al.} in prep.)
No trend with any of these longitudes is evident in these data.}
\label{rholon}
\end{figure}

In addition to the systematic difference between the F ring and the Encke Gap ringlets,
the finite widths of the $\langle\rho\rangle$ distributions in Figure~\ref{rhodist} also suggest that 
there are significant spectral and particle-size variations within each ring. Further evidence for such 
variations can be found in Figures~\ref{rhotrend} and~\ref{ocettrend}, which show 
single-sample estimates of $\rho$ versus optical depth derived from occultations of the star $o$ Ceti,
which were obtained at a very low ring opening angle of 3.5$^\circ$ and hence provide exceptionally
high signal-to-noise data for these rings. In both the Encke Gap ringlet and the F ring
data there are clear trends of increasing $\rho$ with increasing optical depth. The 
denser parts of these rings therefore have a smaller fraction of 1-10 micron 
particles, which would suggest that smaller grains are more widely dispersed in these
rings than the larger ones.  Note that in spite of these trends, Figure~\ref{rhotrend}
demonstrates that systematic differences between the Encke Gap and F ring
persist even at the level of single-sample estimates of $\rho$.

However, while all the F-ring cuts show roughly similar 
rising trends, it is also clear that the detailed structure of the $\rho-\tau$ relationship
differs from occultation to occultation (Figure~\ref{ocettrend}). The optical depth of the ring therefore cannot
be  the only factor controlling its spectral properties. This is consistent
with the observation that occultation cuts of the F ring with similar equivalent depths
can have a finite spread in $\langle\rho\rangle$ values  (see Figure~\ref{rhoscat}, a similar
spread is found if the $\langle\rho\rangle$ values are plotted versus maximum optical depth). 
This scatter in $\langle\rho\rangle$ is not obviously correlated
with longitude relative to the F-ring's pericenter or node, nor does it  seem
to be tied to the moon Prometheus (see Figure~\ref{rholon}). Thus this spectral
variability seems to be independent of the broad-scale structure of the ring.
Instead, the particle size distribution in the F-ring is probably varying on the
much smaller scales associated with the various clumps, strands, fans and 
other features that have been noted in Cassini images \citep{Murray08, Beurle10}.

\section{Spectrally-distinct compact structures in the F ring}

\begin{figure}
\centerline{\resizebox{5in}{!}{\includegraphics{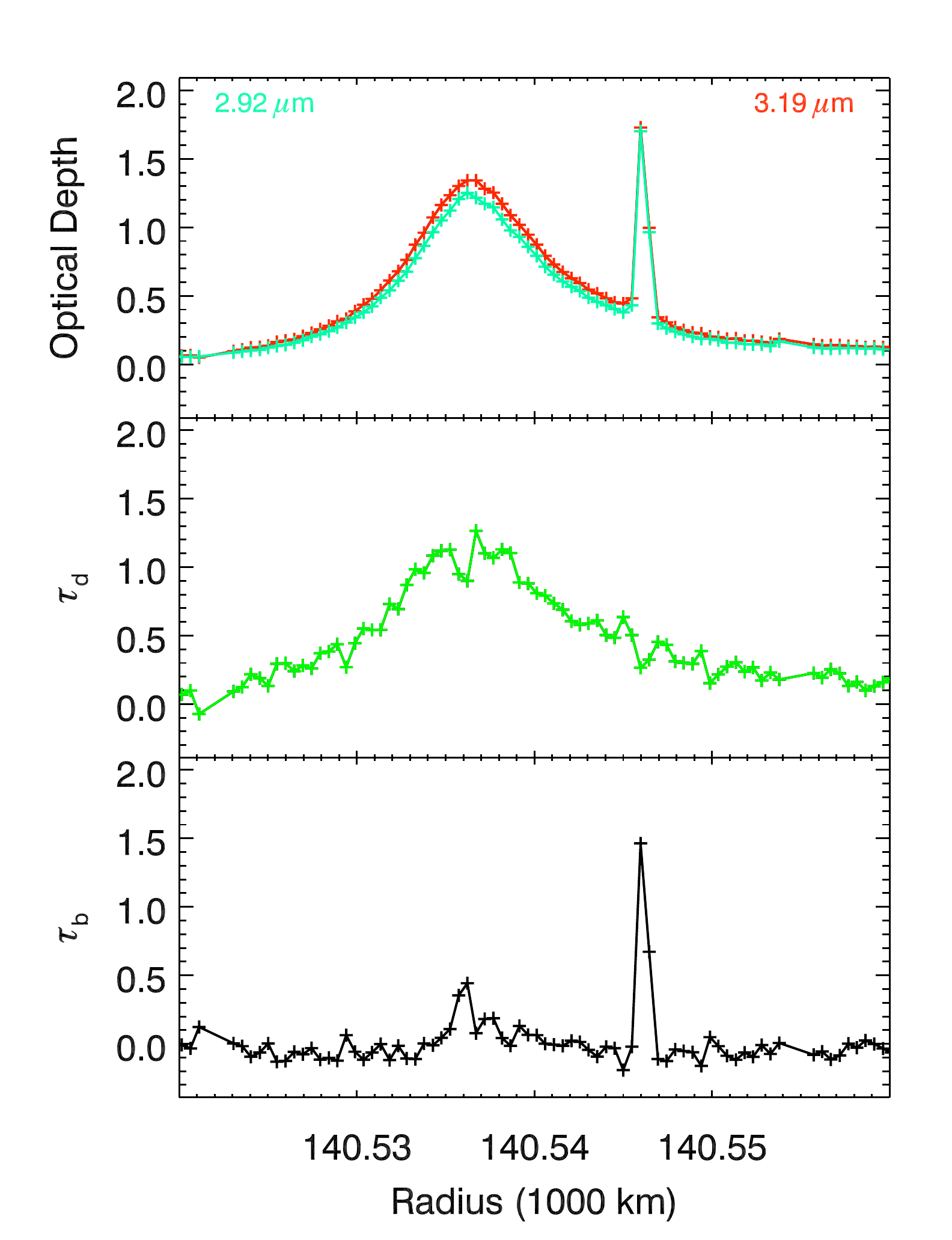}}}
\caption{Optical depth profiles from the Rev 013 egress $\alpha$ Scorpii 
occultation. The top panel shows the normal optical depth profiles at two wavelengths.
Note that the broad peak shows a clear difference in its maximum opacity between the 
two curves, while the sharp peak at 140,546 km does not. The middle panel shows the inferred
optical depth in dust, while the bottom panel shows the inferred optical depth
in excess large particles (see text). The broad peak is  prominent in the former, while
the narrow spike is seen only in the latter, demonstrating that these two features
have very different particle size distributions.}
\label{alpscoplot}
\end{figure}

Interpreting the F-ring's spectral variations is challenging because of
the complex and time-variable nature of the F-ring's morphology,
which complicates efforts to correlate features observed at different times.
However, there are certain features that are sufficiently compact 
and spectrally distinct to clearly stand out from the rest of the F ring. 
One of these features is illustrated in the top panel of Figure~\ref{alpscoplot},
which shows the optical depth profile of the F ring obtained during 
egress of the Rev 013 $\alpha$ Scorpii occultation. This
profile clearly shows two peaks, a broad one at 140,536 km and a narrow one
at 140,546 km. Both of these
peaks were also seen in the simultaneous UVIS occultation trace, which demonstrates
that the narrow feature is a real ring structure and not an instrumental artifact 
\citep{Esposito08, Meinke10}. Comparing the optical depth profiles at 2.9 and 3.2 $\mu$m, we
note that the broader peak is clearly less opaque at 2.9 $\mu$m than it
is at 3.2 $\mu$m, consistent with the ``typical'' F-ring, while  the sharper peak 
has approximately the same opacity in both wavelength channels. 
This suggests that the narrow peak contains a much smaller fraction of 
micron-sized particles than the rest of the F ring.

Narrow optical-depth spikes like the one seen in this profile are
particularly interesting because they represent the same sorts
of compact objects that have been observed
in both imaging \citep{Murray08, Beurle10} and multiple 
UVIS stellar occultations \citep{Esposito08}, and are attributed to 
either small moons or more transient clumps
of debris. Studying these objects in detail has been difficult because
they are embedded in a background of fine dust with complex, time-variable 
structure. This makes quantifying the size of these objects
based on their brightness or opacity alone problematic, because
the signal from the object itself cannot easily be distinguished
from the signal from the surrounding dust. Some of these
objects could even be nothing more than unusually dense clouds of dust,
although others do appear to have sufficient mass to perturb nearby
ring material \citep{Beurle10}. The distinctive transmission spectra
of these objects in VIMS occultations may therefore provide
useful new insights into these structures.

In order to better quantify the distinctive characteristics of the narrow
peak in Figure~\ref{alpscoplot},
consider the following two-component ansatz for the particle size 
distribution in the F ring: On the one hand, there is the ``dust component"
of the ring, consisting of particles smaller than $\sim30$ $\mu$m,
and on the other hand, there is a ``big particle component'' consisting
exclusively of particles larger than $\sim30$ $\mu$m. For this simple
model, we will ignore the variations in the dust size distribution
within the F ring discussed above and assume that the ratio of optical depths for the
dust component  is a constant $\rho_d=0.9$ throughout the ring,
where the specific value is chosen to be close to the peak
of the distribution shown in Figure~\ref{rhodist}. Similarly,
we will assume the big particle component of the ring
has an optical depth ratio $\rho_b=1.0$. In this model,
radial variations in the optical depth ratio $\rho$ are interpreted
as variations in the relative amounts of  ``dust'' and ``big particles''
in different parts of the rings. More specifically, a sharp
increase in $\rho$ indicates an excess
of particles larger than tens of microns across in that particular region
of the ring.  Let us denote the continuum (i.e., 3.2 $\mu$m)
optical depths of these two components as $\tau_d$ and $\tau_b$, 
respectively. Then the slant optical depths at our two standard wavelengths are given by:
\begin{equation}
\tau_{2.9}=\rho_d\tau_d+\tau_b
\end{equation}
and
\begin{equation}
\tau_{3.2}=\tau_d+\tau_b
\end{equation}
Solving these equations for $\tau_d$ and $\tau_b$, 
we find:
\begin{equation}
\tau_{d}=\frac{1}{1-\rho_d}(\tau_{3.2}-\tau_{2.9})
\end{equation}
\begin{equation}
\tau_{b}=\frac{1}{1-\rho_d}(\tau_{2.9}-\rho_d\tau_{3.2})
\end{equation}

Figure~\ref{alpscoplot} shows profiles of $\tau_d$ and $\tau_b$ derived
from the Rev 13 egress $\alpha$ Scorpii occultation using this method.
Note that since $1-\rho_d=0.1$, the noise levels in these profiles are roughly 10
times larger than those of the individual spectral channels.
Nevertheless, this decomposition clearly isolates the narrow spike
from the rest of the F ring. Hence, even if the above model is
a gross oversimplification of the real particle size distribution
in the F ring, it provides a useful method of identifying these
highly distinctive features, which appear to contain higher
concentrations of large particles. It also confirms that at least
this particular feature is significantly less dusty than other parts of
the F ring.

Based on the above findings, we conducted a comprehensive search
through the VIMS occultation data for other spectrally-identifiable
compact structures. This search was done using an automated
routine in order to minimize any subjective bias in feature 
identification. This algorithm is designed to find the most 
prominent spikes in $\tau_b$ profiles and uses multiple
criteria to avoid flagging  various artifacts. In particular, 
a cosmic ray strike in the 3.2 $\mu$m channel will cause
a one-sample-wide spike in transmission and dip in opacity
at that wavelength. This opacity drop reduces the apparent 
strength of the 2.87 $\mu$m dip and therefore produces
a positive spike in the computed $\tau_b$. Such artifacts
can be distinguished from real features in the ring because
the corresponding value of $\tau_d$ is strongly negative. 
Also, these features are always only a single sample wide.
(Note that cosmic ray strikes in the 2.9-micron channel produce
negative spikes in $\tau_b$ and therefore cannot be
mistaken for a clump).

To identify statistically significant features in the F-ring,
we first need to quantify the uncertainties in the parameters
$\tau_b$ and $\tau_d$ for each profile. This is done by removing a smoothed version of 
the relevant profiles (smoothing length 100 km) and computing the $rms$ variations of 
the filtered  values of $\tau_b$ and $\tau_d$ in regions outside the rings
(138,000-139,000 km and 141,000-142,000 km). These $rms$ variations
are denoted $\sigma_b^o$ and $\sigma_d^o$ and characterize the uncertainty in
these parameters when the transmission is close to unity. 
Since the measurement uncertainties
are linear in transmission and not in optical depth, the uncertainty in 
$\tau_b$ and $\tau_d$ when the opacity is finite depends on the background optical depth. 
Fortunately, the DN levels for the VIMS occultations are sufficiently high that the effects
of Poisson counting statistics can be neglected, and using 
standard error propagation we can estimate the $rms$ noise
levels in $\tau_b$ and $\tau_d$ as:
\begin{equation}
\sigma_{b,d}=\sigma^o_{b,d}/T_{3.2}
\end{equation}
where $T_{3.2}$ is the observed transmission of the rings at 3.2 $\mu$m.
 
A real spectrally-distinct region in the ring will have a significantly
positive value of $\tau_b$, and should not have a significantly negative value for
$\tau_d$. Also, the region will only be spectrally identifiable if the 
transmission is not so low that any spectral feature would be saturated.
Finally, features only one sample wide are more likely to be instrumental
or noise artifacts than features with a finite width. Based on these 
considerations, we developed the following four criteria to identify whether a
given measurement of $\tau_b$ and $\tau_d$ in an occultation 
represents part of a spectrally distinctive feature in the ring:
\begin{itemize}
\item The ratio $\tau_b/\sigma_b$ of that sample must be greater than a threshold value 
$S$.
\item The average ratio $\tau_b/\sigma_b$ of that sample and its two nearest
neighbors must be greater than $S/\sqrt{3}$.
\item The ratio $\tau_d/\sigma_d$ of that sample must be greater than -1.
\item The transmission in that sample must be greater than 0.1 (i.e.
the optical depth $\tau_{3.2}$ must be less than 2.3).
\end{itemize}
The first criterion selects the most significant features in the $\tau_b$ profiles,
while the second and third are used to reject noise spikes and cosmic rays.
The last criterion ensures that we have sufficient signal-to-noise to discern
spectral variations. Note that this cut explicitly removes the most opaque
features in the ring, so this search algorithm will not be able to identify 
spectrally-distinct features in the core of the F ring in the roughly dozen
occultations where the peak optical depth exceeds 2.3. However, as we will
see below this algorithm was able to identify one feature with a peak optical
depth of over 2.5 because samples adjacent to the saturated one showed
the required spectral signature. Beyond the second criterion's requirement that
adjacent samples must show evidence of a coherent structure, these criteria 
make no {\it a priori} assumption  about the radial extent of these features.
 
Table~\ref{clumptab} lists the 14 features identified with this technique using
a threshold value $S=3$. Relaxing this threshold significance reveals
additional features that may in fact be clumps, but also admits some structures
that visually appear to be no more than statistical noise in regions
where the total optical depth is high. Thus for the purposes
of this analysis we will focus exclusively on these 14 highest signal-to-noise
features. 

 \begin{figure}
\resizebox{6in}{!}{\includegraphics{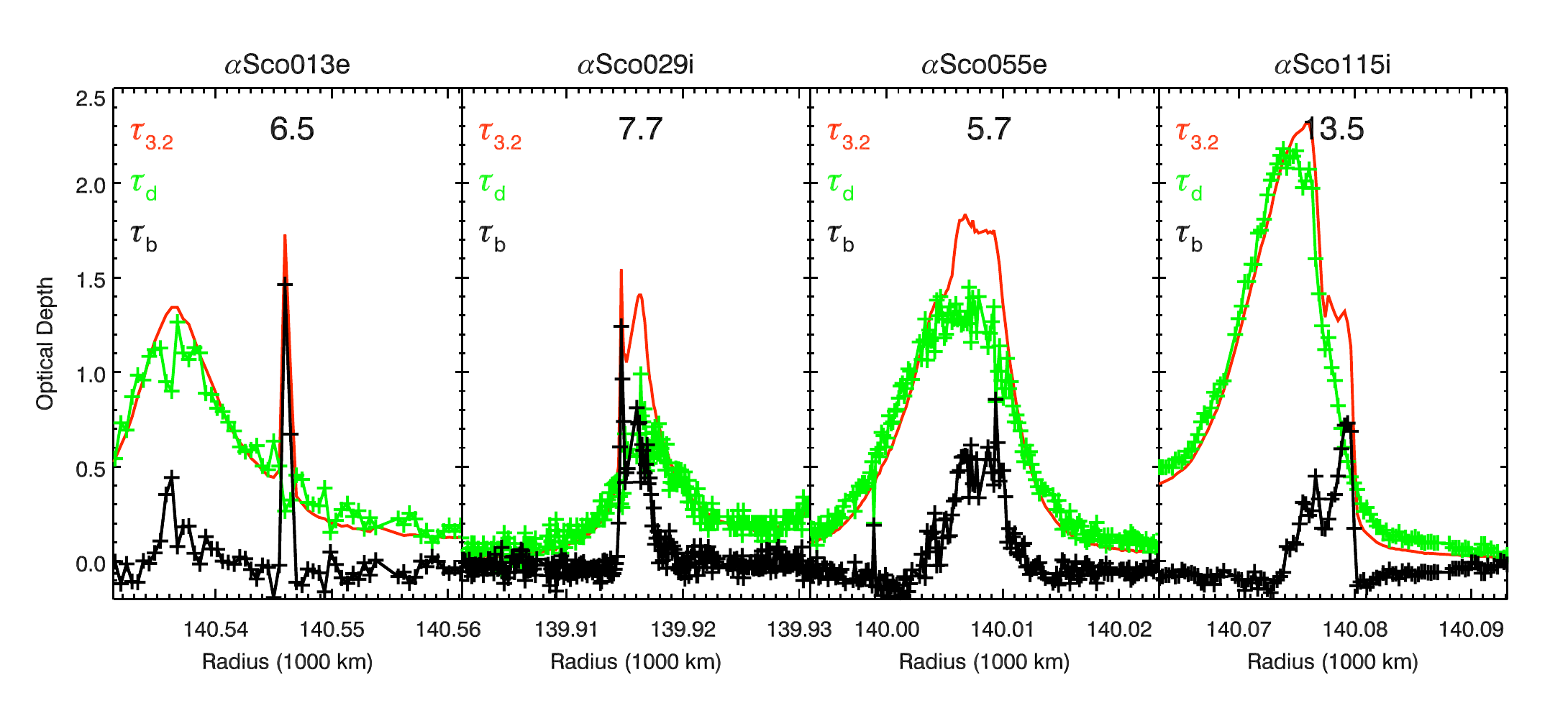}}
\resizebox{6in}{!}{\includegraphics{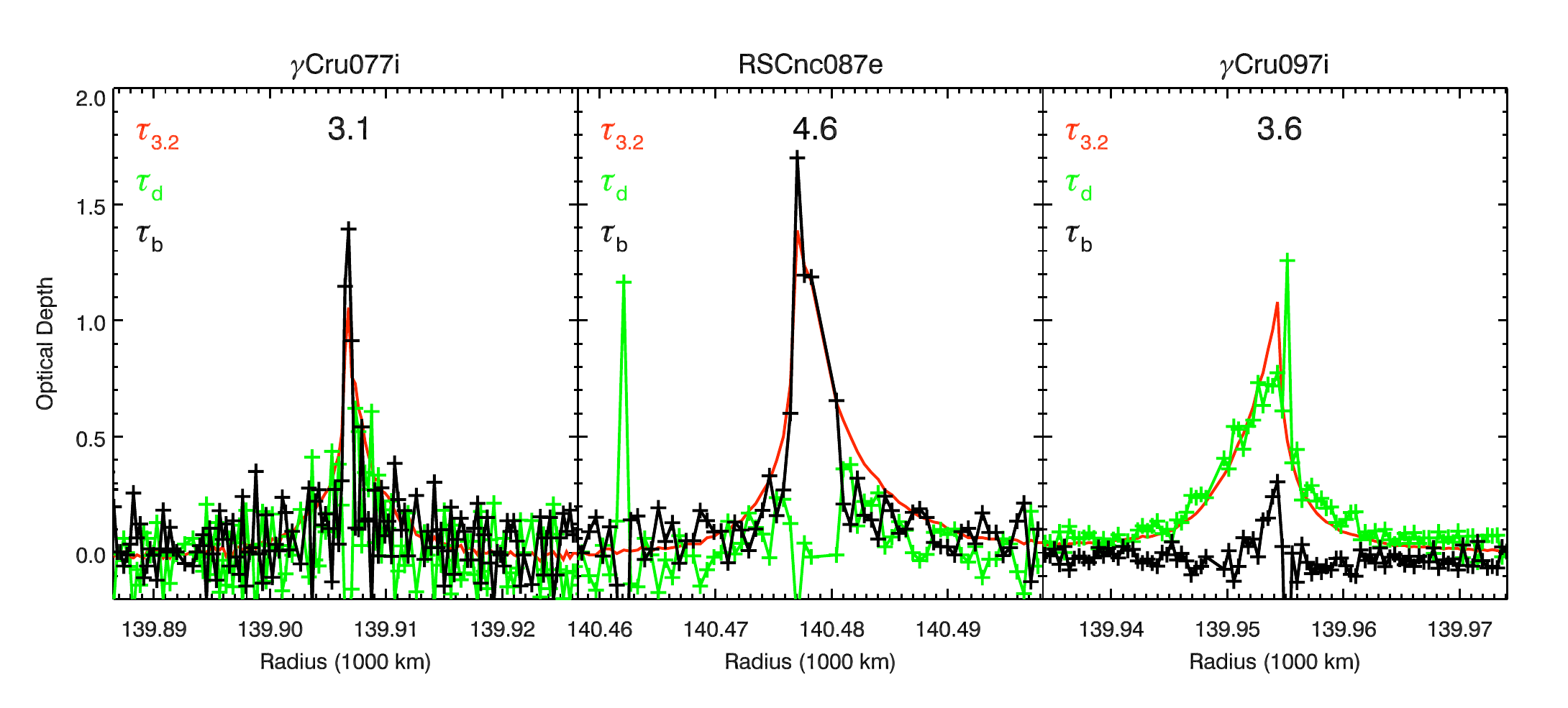}}
\caption{Profiles of the $S>3$ spectrally-distinct clumps identified 
by our automated search. In each plot, the smooth red curve is the
measured optical depth, and the black and green curves are
$\tau_b$ and $\tau_d$, respectively. The numbers on the plot give
the peak $S$ value for each feature. The $\gamma$Cru 77 ingress
data has been down-sampled by a factor of two for clarity. Note the
$\alpha$Sco 115 feature is one of two clumps seen in this
occultation, the other of which is found in a peripheral F-ring strand 
and is shown in Figure~\ref{clumpplot2}.}
\label{clumpplot}
\end{figure}

\begin{figure}
\resizebox{6in}{!}{\includegraphics{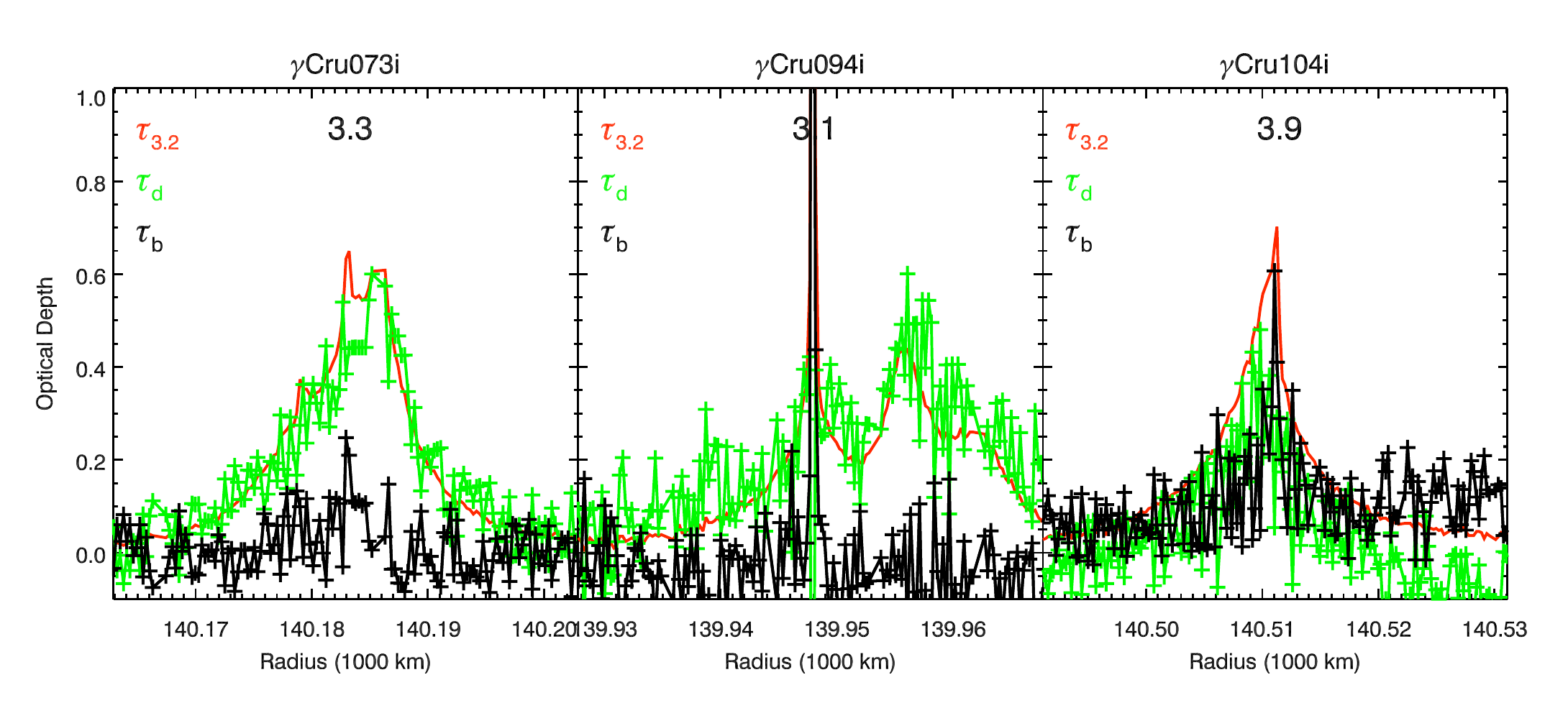}}
\resizebox{6in}{!}{\includegraphics{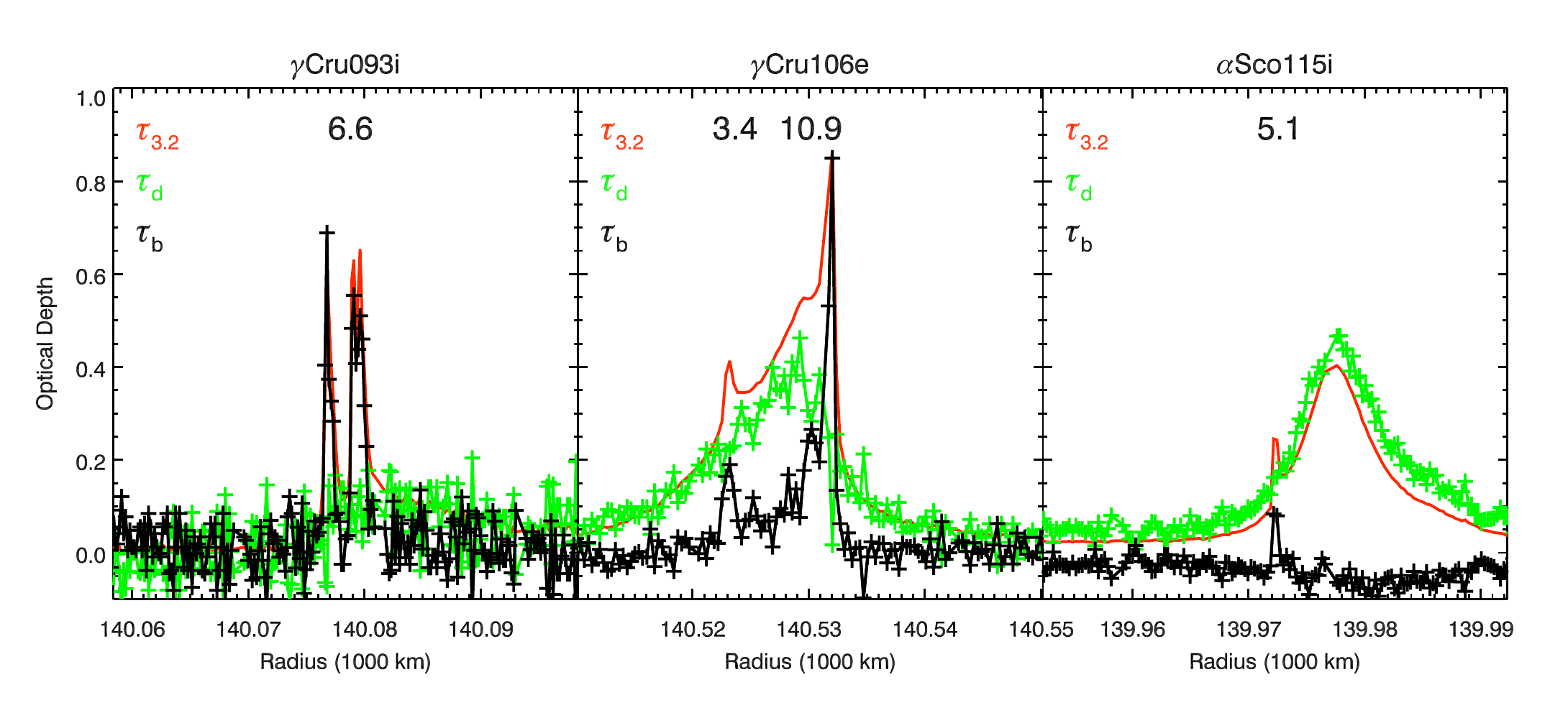}}
\caption{Profiles of the $S>3$ spectrally-distinct clumps identified 
by our automated search. In each plot, the smooth red curve is the
measured optical depth, and the black and green curves are
$\tau_b$ and $\tau_d$, respectively. The numbers on the plot give
the peak $S$ value for each feature. The $\gamma$Cru 094, 104 
and 106 data have been down-sampled by a factor of two for clarity.
Note the $\alpha$Sco 115 feature is one of two clumps seen in this
occultation, the other of which is found in the core of the F ring and is 
shown in Figure~\ref{clumpplot}.}
\label{clumpplot2}
\end{figure}

Figures~\ref{clumpplot} and~\ref{clumpplot2} illustrate the radial profiles
of these features.  In each case, the $\tau_b$ profile shows a relatively
compact structure less than 10 km wide, while the $\tau_d$ profile is
fairly smooth (even in cases like the Rev 106 $\gamma$ Crucis occultation
where the optical depth profile contains multiple peaks), which confirms
that spectral decomposition works sensibly in all of these cases, 
despite the variations in the background $\rho$ of the dust.
While many features correspond to obvious  spikes in
the optical depth profiles and therefore could have been identified 
based on their morphology alone (cf. Esposito {\it et al} 2008),  several
of these features appear as rather subtle features in the
optical depth profile and would be difficult to identify
if not for their unusual spectral properties. 

None of these features is opaque (the highest observed 
optical depth associated with these features being 2.5), so
none of these features can be ascribed to a resolved moonlet
in the F ring similar to that found in the $\alpha$ Leonis UVIS occultation
\citep{Esposito08}. Instead, what we appear to be seeing are relatively
compact clumps of debris with enhanced large particle populations.

The morphologies of these debris clouds are quite diverse, and include:
\begin{itemize}
\item
Isolated sub-kilometer-wide spikes ($\alpha$Sco013, 
$\gamma$Cru073,  $\gamma$Cru094, 
and $\alpha$Sco115 at 139,972 km).
\item 
Clusters of multiple narrow spikes ($\gamma$Cru093).
\item
Broader features several kilometers wide ($\alpha$Sco055, 
RSCnc087,  and $\alpha$Sco115 at 140,078 km).
\item 
Combinations of sub-kilometer spikes with more diffuse features.
($\alpha$Sco029 and $\gamma$Cru106)
\end{itemize}
The latter two types of features only become obvious through the 
spectral decomposition, and reveal that these clumps can have
significant substructure. Note the signal-to-noise of the 
$\gamma$Cru077, 097 and 104 occultations are low, 
so the morphologies of these three features are 
difficult to determine.


\begin{figure}
\resizebox{6in}{!}{\includegraphics{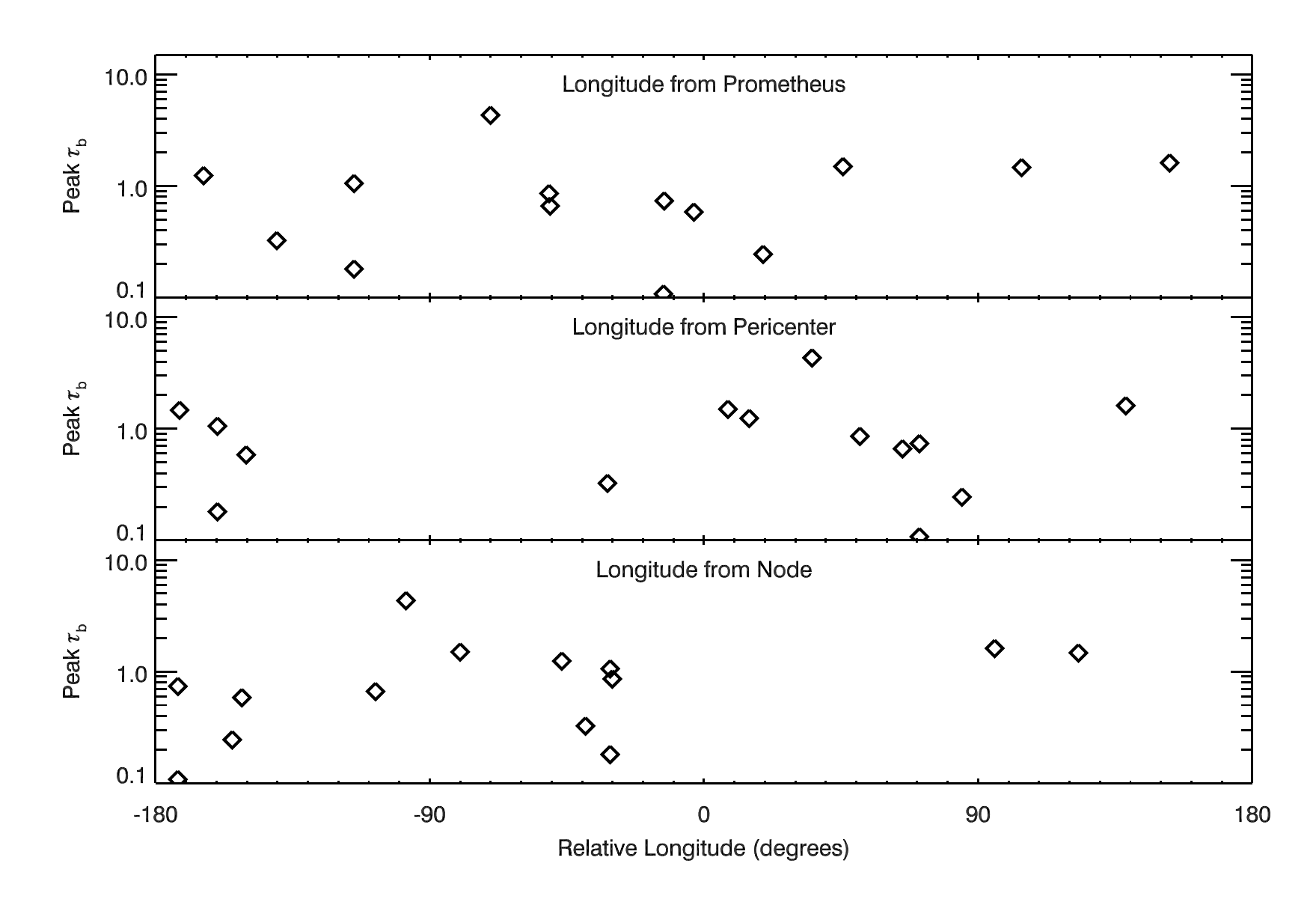}}
\caption{The distribution of spectrally-identifiable regions in the F ring
listed in Table~\ref{clumptab} in optical depth and longitude 
relative to Prometheus and the
F-ring's pericenter and node.}
\label{clumplon}
\end{figure}

Some have argued that these clumps in the F ring arise
from accretion of material initiated by the orbital perturbations
produced by close encounters with Prometheus \citep{Beurle10}, and
there have even been attempts to demonstrate this connection with
Prometheus by looking for clustering of these features in 
longitude relative to Prometheus \citep{Meinke10}. Figure~\ref{clumplon} 
shows the distribution of the high-quality spectrally-identifiable
features in longitude relative to Prometheus. We find no evidence
for clustering in these data. This figure also shows the distribution
of these features in longitude relative to the F-ring's pericenter and node.
Here there may be some hints of clustering,  but
the evidence is weak. These data
do not provide convincing support for the idea that these particular clumps
are organized in any coherent way on large spatial scales
in the F ring. However, the absence of this particular observational
signature does not rule out the possibility that Prometheus is
responsible for at least some of these F-ring features. Perturbations
in the F-ring induced by Prometheus close encounters can sometimes
persist for more than one synodic period \citep{Murray08, Beurle10}, and if these
knots correspond to the locations of these spectrally-identifiable features,
we might expect their distribution to be roughly uniform in longitude.

\section{Discussion}

The variations in the transmission spectra described above can be
divided into two types: those that appear to be correlated
with optical depth (such as the trends seen in Figures~\ref{rhotrend}
and~\ref{ocettrend}) and those that are not (such as the 
systematic spectral differences between the Encke Gap ringlets 
and the F ring in Figure~\ref{rhoscat},  or between the spectrally-distinctive clumps and the rest of the F ring in Figure~\ref{alpscoplot}). While all of these spectral variations must 
reflect differences in the local particle size distribution, the physical 
processes responsible for the trends with optical depth are
likely very different from those that distinguish the Encke Gap ringlets
and F-ring clumps from the typical F ring and Charming Ringlet. Thus 
we will consider these two phenomena separately below.

\subsection{Spectral variations correlated with the rings' optical depth}

In both the F ring and the Encke Gap ringlets, the
dip in the extinction spectra tends to be stronger 
in regions of lower optical depth (see Figures~\ref{rhotrend} and~\ref{ocettrend}),
so the fraction of particles in the 1-10 micron size range
increases as the optical depth decreases. Larger ring particles
therefore appear to be more concentrated in the denser parts 
of the rings, while the smaller particles seem to be more widely 
dispersed in space. 
There are multiple physical processes that could potentially lead to this
particle stratification. For example, any non-gravitational forces that
might be perturbing these particles' orbits would tend to produce larger accelerations
with smaller grains, dispersing them over a wider area of phase space. Alternatively,
under certain conditions particle-particle interactions such as physical collisions or even Coulomb scattering could affect the orbital elements of small particles more than larger particles, and thus produce a more dispersed population of small grains.
Detailed investigations of these phenomena (which are beyond the scope of this work) should therefore lead to a better understanding of these rings' local environment and the particles' charge states and surface properties.



\subsection{Spectral variations not correlated with the rings' optical depth}

The two most striking spectral variations that are not correlated
with the rings' optical depth are (1) the systematically weaker opacity dips
in the Encke Gap ringlets compared to the Charming Ringlet and
typical F ring (cf. Figure~\ref{rhoscat} and~\ref{rhotrend}), and (2) the strongly attenuated opacity dips in certain 
compact regions within the F ring (cf. Figure~\ref{alpscoplot}). These spectral features indicate
that the Encke Gap ringlets and certain compact features in the F ring
have smaller fractions of few-micron-sized particles than regions of comparable optical depth in the typical F ring and Charming Ringlet.  Such differences
in the local particle size distribution are likely due to some change in
the balance between dust production, accretion, fragmentation and loss.
Given that multiple aspects of these rings' dynamical environment
influence these processes, a complete investigation of these features
is well beyond the scope of this report. Instead, we will briefly consider one 
particular phenomenon that may be relevant: variations in the rings'  local velocity dispersion altering the outcome of collisions among ring particles.

Despite the low overall optical depth of these rings, collisions among
particles within the ring are not extremely rare. The relevant collision 
rate is proportional to the product of the particles'  orbital mean motion and the
ring's local optical depth. While the typical optical depths of these rings are rather low ($<0.1$), the orbital frequencies are sufficiently high ($>1/$day) 
that the timescale for collisions is rather short, being only weeks or 
less in the typical F ring. These collisions can have a significant
impact on the ring's particle size distribution because, depending on how fast 
particles of different sizes approach each other, they can either stick together, bounce apart or  fragment into pieces.

\citet{Guttler10} provide detailed maps of collision outcomes
versus particle size and impact speeds for non-icy particles, 
which show that the transitions
between sticking and bouncing for silica-rich dust grains and aggregates 
are size-dependent and typically fall in the range between 0.01 and 100 cm/s. 
While the precise locations of these transitions will likely be different 
for the ice-rich ring particles, elementary theoretical calculations
indicate that the relevant threshold speeds are of a similar
order of magnitude. In particular, 
the threshold speed separating bouncing from  simple sticking 
(without compaction) should occur when the impact energy equals  
$5E_{roll}$, where $E_{roll}$ is the energy
dissipated when two particles roll over each other by 90$^\circ$, which
depends on the material content and size of the grains \citep{DT97,Guttler10}. 
The corresponding impact speed for identically-sized grains of mass $m$ and
radius $R$ can be written as:
\begin{equation}
v_{crit}=\sqrt{\frac{5E_{roll}}{m}}=\sqrt{\frac{45\pi\gamma\xi_{crit}}{4d R^2}}
\end{equation}
where $d$ is the mass density of the particles, $\gamma$ is the interface energy
and $\xi_{crit}$ is a critical rolling displacement. For ice, these numbers have been 
approximated as
$d=1$ g/cm$^3$, $\gamma \sim 100$ erg/cm$^2$ and $\xi_{crit} \sim$ 8 Angstroms
\citep{Wada09}. Inserting these values in the above expression, we find
$v_{crit} \sim 20$ cm/s $(10 \mu$m/$R$). Thus particles smaller than 10 $\mu$m across
could begin sticking to other particles whenever the collision velocities fall below 20 cm/s.
Of course, other processes such as the compaction of loose aggregates 
could potentially complicate the situation, but even so $v_{crit}$ should still
provide a useful fiducial impact speed below which particles are more
apt to stick together. Thus, any dynamical process that reduces
the local  velocity dispersion within these rings much below 
$\sim 20$ cm/s could lead to a depletion of small grains. 

Recent simulations of the interactions between Prometheus and the F ring
have demonstrated that the moon's perturbations on the ring particle's orbits
can produce localized regions of enhanced density and reduced velocity
dispersions \citep{Beurle10}. The lowest velocity dispersions observed in these
simulations are of order 2 cm/s, well below the critical speed for simple sticking
with 10 $\mu$m grains. The spectrally-distinct compact regions in the F ring could
therefore be interpreted as regions 
where interactions with Prometheus have increased the local particle density and
decreased the local velocity dispersion, making it more likely that  
the 1-10 $\mu$m grains would have become attached to larger grains. 
The lack of correlation between the compact regions and longitude relative
to Prometheus is not a major problem for this model if the disturbed regions persist for multiple synodic cycles and thus are roughly equally likely to be
found anywhere around the ring.   This interpretation would also 
be consistent with the emerging idea that some compact structures
in the F ring could represent the accretion of material within the rings \citep{Beurle10,
Meinke10}. In this case,
the VIMS data would provide the first direct evidence for small
grains adhering together to form larger aggregates.

For the systematically low fraction of 1-10 $\mu$m sized grains in the Encke Gap ringlets, the relevance of such moon-induced aggregation is far less clear. While the particles in all three Encke Gap ringlets periodically encounter
the moon Pan either as they drift by the moon (for the inner and outer ringlets) or undergo horseshoe motion (for the central ringlet), it is not obvious whether these encounters could produce regions of reduced velocity dispersions like
those predicted for the F ring. In the F ring, regions  of reduced velocity dispersion and increased density  arise due
to the longitudinal variations in the perturbations to the ring 
particles' semi-major axes, which arise because of the epicyclic motion
of Prometheus and the F ring relative to each other \citep{Beurle10}. 
Since Pan is on  a nearly circular orbit, similar longitudinal gradients 
in the semi-major axis shift cannot be produced within the Encke Gap by Pan's epicyclic motion. However, the Encke Gap ringlets, like the Charming Ringlet, exhibit
some degree of heliotropic behavior, in which solar radiation pressure 
causes the geometrical center of the ringlet to be displaced from the 
center of Saturn towards the sun \citep{Hedman07H, Hedman10C}. 
Due to this forced eccentricity in the ring particles' orbits, particles encountering Pan
at different longitudes relative to the sun will be at different phases in their 
epicyclic motion. This should lead to longitudinal variations in the semi-major axis 
perturbations like those required to produce regions of enhanced density
and reduced velocity dispersion. Such a model could potentially explain why the Encke Gap ringlets have distinctly different size distributions than the
Charming ringlet, whose gap is devoid of Pan-sized moons. 
However, a possible problem with this idea is that the Pan-induced perturbations on the different Encke ringlets have different strengths and occur on
different timescales, so it is not clear if such perturbations would
affect the size distributions of all three ringlets equally. 

Of course, there are a number of other possible
processes that might also be responsible for the observed
variations in these rings'  particle size 
distributons.  For example, the overall 
dynamical environments the rings  inhabit may influence how
 efficiently particles of different sizes are generated and how quickly particles
are lost to the system by collisions with nearby rings and moons. 
In this context, the compact F-ring structures might
correspond to regions where larger particles are produced more rapidly than 
other parts of the ring, perhaps again due to differences in the
typical impact speeds into and between particles. Future detailed
comparisons of these spectral data with predictions derived from
various models of the revelant processes should yield 
useful new insights into the dynamics
and particle properties of these dusty rings.

\medskip

{\bf Acknowledgments:} We wish to acknowledge NASA, the Cassini project and
the VIMS team for providing the data that made this analysis possible.
We also thank H. Throop and J. N. Cuzzi for their detailed and insightful
reviews of this manuscript, and L. Esposito and C. Murray for the useful discussions.
This work was supported in part by a Cassini Data Analysis Program grant NNX09AE74G.


\begin{thebibliography}{}

\bibitem[{Arnott} {\em et~al.}(1995){Arnott}, {Dong}, and {Hallett}]{Arnott95}
{Arnott}, W.~P., Y.~Y. {Dong},\ and J.~{Hallett} 1995.
\newblock {Extinction efficiency in the infrared (2-18$\mu$m) of laboratory ice
  clouds: observations of scattering minima in the Christiansen bands of ice}.
\newblock {\em Appl. Opt.\/}~{\bf 34}, 541--551.

\bibitem[{Beurle} {\em et~al.}(2010){Beurle}, {Murray}, {Williams}, {Cooper},
  and {Agnor}]{Beurle10}
{Beurle}, K., C.~D. {Murray}, G.~A. {Williams}, N.~J. {Cooper},\ and C.~B.
  {Agnor} 2010.
\newblock {Direct evidence for gravitational instability and moonlet formation
  in Saturn's rings}.
\newblock {\em ApJL\/}~{\bf 718}, 176--180.

\bibitem[{Bosh} {\em et~al.}(2002){Bosh}, {Olkin}, {French}, and
  {Nicholson}]{Bosh02}
{Bosh}, A.~S., C.~B. {Olkin}, R.~G. {French},\ and P.~D. {Nicholson} 2002.
\newblock {Saturn's F ring: Kinematics and particle sizes from stellar
  occultation studies}.
\newblock {\em Icarus\/}~{\bf 157}, 57--75.

\bibitem[{Brown} {\em et~al.}(2004){Brown}, {Baines}, {Bellucci}, {Bibring},
  {Buratti}, {Capaccioni}, {Cerroni}, {Clark}, {Coradini}, {Cruikshank},
  {Drossart}, {Formisano}, {Jaumann}, {Langevin}, {Matson}, {McCord},
  {Mennella}, {Miller}, {Nelson}, {Nicholson}, {Sicardy}, and {Sotin}]{Brown04}
{Brown}, R.~H., K.~H. {Baines}, G.~{Bellucci}, J.-P. {Bibring}, B.~J.
  {Buratti}, F.~{Capaccioni}, P.~{Cerroni}, R.~N. {Clark}, A.~{Coradini}, D.~P.
  {Cruikshank}, P.~{Drossart}, V.~{Formisano}, R.~{Jaumann}, Y.~{Langevin},
  D.~L. {Matson}, T.~B. {McCord}, V.~{Mennella}, E.~{Miller}, R.~M. {Nelson},
  P.~D. {Nicholson}, B.~{Sicardy},\ and C.~{Sotin} 2004.
\newblock {The Cassini Visual And Infrared Mapping Spectrometer (VIMS)
  Investigation}.
\newblock {\em Space Science Reviews\/}~{\bf 115}, 111--168.

\bibitem[{Christiansen}(1884){Christiansen}]{Chris84}
{Christiansen}, C. 1884.
\newblock {Untersuchungen {\"u}ber die optischen Eigenschaften von fein
  vertheilten K{\"o}rpern}.
\newblock {\em Annalen der Physik und Chemie\/}~{\bf 23}, 298--306.

\bibitem[{Christiansen}(1885){Christiansen}]{Chris85}
{Christiansen}, C. 1885.
\newblock {Untersuchungen {\"u}ber die optischen Eigenschaften von fein
  vertheilten K{\"o}rpern}.
\newblock {\em Annalen der Physik und Chemie\/}~{\bf 24}, 439--446.

\bibitem[{Cuzzi} {\em et~al.}(2009){Cuzzi}, {Clark}, {Filacchione}, {French},
  {Johnson}, {Marouf}, and {Spilker}]{Cuzzi09}
{Cuzzi}, J., R.~{Clark}, G.~{Filacchione}, R.~{French}, R.~{Johnson},
  E.~{Marouf},\ and L.~{Spilker} 2009.
\newblock {Ring Particle Composition and Size Distribution}.
\newblock In {Dougherty, M.~K., Esposito, L.~W., \& Krimigis, S.~M.} (Ed.),
  {\em Saturn from Cassini-Huygens}, pp.\  459--509.

\bibitem[{Cuzzi}(1985){Cuzzi}]{Cuzzi85}
{Cuzzi}, J.~N. 1985.
\newblock {Rings of Uranus - Not so thick, not so black}.
\newblock {\em Icarus\/}~{\bf 63}, 312--316.

\bibitem[{Dominik} and {Tielens}(1997){Dominik} and {Tielens}]{DT97}
{Dominik}, C.,\ and A.~G.~G.~M. {Tielens} 1997.
\newblock {The Physics of Dust Coagulation and the Structure of Dust Aggregates
  in Space}.
\newblock {\em ApJ\/}~{\bf 480}, 647--673.

\bibitem[{Elachi} and {van Zyl}(2006){Elachi} and {van Zyl}]{EvZ06}
{Elachi}, C.,\ and J.~{van Zyl} 2006.
\newblock {\em {Introduction to the Physics and Techniques of Remote Sensing:
  Second Edition}}.
\newblock John Wiley.

\bibitem[{Esposito} {\em et~al.}(2008){Esposito}, {Meinke}, {Colwell},
  {Nicholson}, and {Hedman}]{Esposito08}
{Esposito}, L.~W., B.~K. {Meinke}, J.~E. {Colwell}, P.~D. {Nicholson},\ and
  M.~M. {Hedman} 2008.
\newblock {Moonlets and clumps in Saturn's F ring}.
\newblock {\em Icarus\/}~{\bf 194}, 278--289.

\bibitem[{French} and {Nicholson}(2000){French} and {Nicholson}]{FN00}
{French}, R.~G.,\ and P.~D. {Nicholson} 2000.
\newblock {Saturn's Rings II. Particle sizes inferred from stellar occultation
  data}.
\newblock {\em Icarus\/}~{\bf 145}, 502--523.

\bibitem[{French} {\em et~al.}(1993){French}, {Nicholson}, {Cooke}, {Elliot},
  {Matthews}, {Perkovic}, {Tollestrup}, {Harvey}, {Chanover}, {Clark},
  {Dunham}, {Forrest}, {Harrington}, {Pipher}, {Brahic}, {Grenier}, {Roques},
  and {Arndt}]{French93}
{French}, R.~G., P.~D. {Nicholson}, M.~L. {Cooke}, J.~L. {Elliot},
  K.~{Matthews}, O.~{Perkovic}, E.~{Tollestrup}, P.~{Harvey}, N.~J. {Chanover},
  M.~A. {Clark}, E.~W. {Dunham}, W.~{Forrest}, J.~{Harrington}, J.~{Pipher},
  A.~{Brahic}, I.~{Grenier}, F.~{Roques},\ and M.~{Arndt} 1993.
\newblock {Geometry of the Saturn system from the 3 July 1989 occultation of 28
  SGR and Voyager observations}.
\newblock {\em Icarus\/}~{\bf 103}, 163--214.

\bibitem[{French} {\em et~al.}(1991){French}, {Nicholson}, {Porco}, and
  {Marouf}]{French91}
{French}, R.~G., P.~D. {Nicholson}, C.~C. {Porco},\ and E.~A. {Marouf} 1991.
\newblock {Dynamics and structure of the Uranian rings}.
\newblock In J.~T. {Bergstralh}, E.~D. {Miner}, and M.~S. {Matthews} (Eds.),
  {\em Uranus}, pp.\  327--409.

\bibitem[{G{\"u}ttler} {\em et~al.}(2010){G{\"u}ttler}, {Blum}, {Zsom},
  {Ormel}, and {Dullemond}]{Guttler10}
{G{\"u}ttler}, C., J.~{Blum}, A.~{Zsom}, C.~W. {Ormel},\ and C.~P. {Dullemond}
  2010.
\newblock {The outcome of protoplanetary dust growth: pebbles, boulders, or
  planetesimals?. I. Mapping the zoo of laboratory collision experiments}.
\newblock {\em A\&A\/}~{\bf 513}, A56+.

\bibitem[{Hapke}(1993){Hapke}]{Hapke93}
{Hapke}, B. 1993.
\newblock {\em {Theory of reflectance and emittance spectroscopy}}.
\newblock Cambridge University Press.

\bibitem[{Hedman} {\em et~al.}(2007){Hedman}, {Burns}, {Tiscareno}, and
  {Porco}]{Hedman07H}
{Hedman}, M.~M., J.~A. {Burns}, M.~S. {Tiscareno},\ and C.~C. {Porco} 2007.
\newblock {The Heliotropic Rings of Saturn}.
\newblock In {\em Bulletin of the American Astronomical Society}, Volume~38 of
  {\em Bulletin of the American Astronomical Society}, pp.\  427--+.

\bibitem[{Hedman} {\em et~al.}(2010){Hedman}, {Burt}, {Burns}, and
  {Tiscareno}]{Hedman10C}
{Hedman}, M.~M., J.~A. {Burt}, J.~A. {Burns},\ and M.~S. {Tiscareno} 2010.
\newblock {The shape and dynamics of a heliotropic dusty ringlet in the Cassini
  Division}.
\newblock {\em Icarus\/}~{\bf 210}, 284--297.

\bibitem[{Hedman} {\em et~al.}(2010){Hedman}, {Nicholson}, {Baines}, {Buratti},
  {Sotin}, {Clark}, {Brown}, {French}, and {Marouf}]{Hedman10}
{Hedman}, M.~M., P.~D. {Nicholson}, K.~H. {Baines}, B.~J. {Buratti},
  C.~{Sotin}, R.~N. {Clark}, R.~H. {Brown}, R.~G. {French},\ and E.~A. {Marouf}
  2010.
\newblock {The Architecture of the Cassini Division}.
\newblock {\em AJ\/}~{\bf 139}, 228--251.

\bibitem[{Hedman} {\em et~al.}(2007){Hedman}, {Nicholson}, {Salo}, {Wallis},
  {Buratti}, {Baines}, {Brown}, and {Clark}]{Hedman07}
{Hedman}, M.~M., P.~D. {Nicholson}, H.~{Salo}, B.~D. {Wallis}, B.~J. {Buratti},
  K.~H. {Baines}, R.~H. {Brown},\ and R.~N. {Clark} 2007.
\newblock {Self-gravity wake structures in Saturn's A ring revealed by Cassini
  VIMS}.
\newblock {\em AJ\/}~{\bf 133}, 2624--2629.

\bibitem[{Hor{\'a}nyi} {\em et~al.}(2009){Hor{\'a}nyi}, {Burns}, {Hedman},
  {Jones}, and {Kempf}]{Horanyi09}
{Hor{\'a}nyi}, M., J.~A. {Burns}, M.~M. {Hedman}, G.~H. {Jones},\ and
  S.~{Kempf} 2009.
\newblock {Diffuse Rings}.
\newblock In {Dougherty, M.~K., Esposito, L.~W., \& Krimigis, S.~M.} (Ed.),
  {\em Saturn from Cassini-Huygens}, pp.\  511--536. Springer.

\bibitem[{Liou} {\em et~al.}(1998){Liou}, {Yang}, {Takano}, {Sassen},
  {Charlock}, and {Arnott}]{Liou98}
{Liou}, K.~N., P.~{Yang}, Y.~{Takano}, K.~{Sassen}, T.~{Charlock},\ and
  W.~{Arnott} 1998.
\newblock {On the radiative properties of contrail cirrus}.
\newblock {\em GRL\/}~{\bf 25}, 1161--1164.

\bibitem[{Mastrapa} {\em et~al.}(2009){Mastrapa}, {Sandford}, {Roush},
  {Cruikshank}, and {Dalle Ore}]{Mastrapa09}
{Mastrapa}, R.~M., S.~A. {Sandford}, T.~L. {Roush}, D.~P. {Cruikshank},\ and
  C.~M. {Dalle Ore} 2009.
\newblock {Optical Constants of Amorphous and Crystalline H$_{2}$O-ice: 2.5-22
  $\mu$m (4000-455 cm$^{-1}$) Optical Constants of H$_{2}$O-ice}.
\newblock {\em ApJ\/}~{\bf 701}, 1347--1356.

\bibitem[{Meinke} {\em et~al.}(2010){Meinke}, {Esposito}, and
  {Albers}]{Meinke10}
{Meinke}, B.~K., L.~W. {Esposito},\ and N.~{Albers} 2010.
\newblock {UVIS ring occultations show F ring feature location and optical
  depth correlated with Prometheus}.
\newblock In {\em Bulletin of the American Astronomical Society}, Volume~41 of
  {\em Bulletin of the American Astronomical Society}, pp.\  939.

\bibitem[{Murray} {\em et~al.}(2008){Murray}, {Beurle}, {Cooper}, {Evans},
  {Williams}, and {Charnoz}]{Murray08}
{Murray}, C.~D., K.~{Beurle}, N.~J. {Cooper}, M.~W. {Evans}, G.~A. {Williams},\
  and S.~{Charnoz} 2008.
\newblock {The determination of the structure of Saturn's F ring by nearby
  moonlets}.
\newblock {\em Nature\/}~{\bf 453}, 739--744.

\bibitem[{Nicholson} and {Hedman}(2010){Nicholson} and {Hedman}]{NH10}
{Nicholson}, P.~D.,\ and M.~M. {Hedman} 2010.
\newblock {Self-gravity wake parameters in Saturn's A and B rings}.
\newblock {\em Icarus\/}~{\bf 206}, 410--423.

\bibitem[{Prost}(1968){Prost}]{Prost68}
{Prost}, R. 1968.
\newblock {The influence of the Christiansen Effect on I.R. spectra of
  powders}.
\newblock {\em Clays and Clay Minerals\/}~{\bf 21}, 363--368.

\bibitem[{Showalter} {\em et~al.}(1992){Showalter}, {Pollack}, {Ockert},
  {Doyle}, and {Dalton}]{Showalter92}
{Showalter}, M.~R., J.~B. {Pollack}, M.~E. {Ockert}, L.~R. {Doyle},\ and J.~B.
  {Dalton} 1992.
\newblock {A photometric study of Saturn's F Ring}.
\newblock {\em Icarus\/}~{\bf 100}, 394--411.

\bibitem[{Vahidinia} {\em et~al.}(2011){Vahidinia}, {Cuzzi}, {Hedman},
  {Draine}, {Clark}, {Roush}, {Filacchione}, {Nicholson}, {Brown}, {Buratti},
  and {Sotin}]{Vahidinia11}
{Vahidinia}, S., J.~{Cuzzi}, M.~{Hedman}, B.~{Draine}, R.~{Clark}, T.~{Roush},
  G.~{Filacchione}, P.~{Nicholson}, R.~{Brown}, B.~{Buratti},\ and C.~{Sotin}
  2011.
\newblock {Saturn's F ring grains: aggregates of crystalline water ice}.
\newblock {\em submitted to Icarus\/}, --.

\bibitem[{van de Hulst}(1957){van de Hulst}]{vandeHulst}
{van de Hulst}, H.~C. 1957.
\newblock {\em {Light Scattering by Small Particles}}.
\newblock Light Scattering by Small Particles, New York: John Wiley and Sons,
  1957.

\bibitem[{Wada} {\em et~al.}(2009){Wada}, {Tanaka}, {Suyama}, {Kimura}, and
  {Yamamoto}]{Wada09}
{Wada}, K., H.~{Tanaka}, T.~{Suyama}, H.~{Kimura},\ and T.~{Yamamoto} 2009.
\newblock {Collisional Growth Conditions for Dust Aggregates}.
\newblock {\em ApJ\/}~{\bf 702}, 1490--1501.

\end{thebibliography}

\pagebreak

 
\begin{table}
\caption{Stellar occulations of the F ring}
\label{obstab}
\centerline{\resizebox{6in}{!}{\begin{tabular}{|l c c | c | c | c c | c c |c c|c c|c|}\hline 
Star & Rev & ingress/ & Sum$^a$ & Date$^b$ & B$^c$ & Longitude$^d$ & Max DN$^e$ & Max DN$^e$ & Min T$^f$ & Min T$^f$ & Max $\tau$ & Max $\tau$ & $<{\rho}>^g$ \\ 
 & & egress & & & (deg) & (deg) & 2.92 $\mu$m & 3.19$\mu$m & 2.92 $\mu$m & 3.19$\mu$m & 
 2.92 $\mu$m & 3.19$\mu$m &\\
 \hline
$o$Cet&008&i & S& 2005-144T05:00 &  3.45 &   20.12 &  994. & 1181. & 0.04 & 0.03 & 3.34 & 3.57 & 0.85 \\
$o$Cet&008&e & S& 2005-144T08:04 &  3.45 &  -47.08 &  995. & 1178. & 0.03 & 0.02 & 3.43 & 3.92 & 0.80 \\
$o$Cet&009&i & S& 2005-162T08:06 &  3.45 &   10.75 &  990. & 1164. & 0.01 & 0.00 & 4.95 & 5.34 & 0.81 \\
$o$Cet&009&e & S& 2005-162T10:25 &  3.45 &  -37.57 &  998. & 1166. & 0.11 & 0.08 & 2.23 & 2.57 & 0.82 \\
$o$Cet&010&i & S& 2005-180T12:39 &  3.45 &    3.58 & 1184. & 1370. & 0.00 & 0.00 & 5.43 & 5.39 & 0.83 \\
$o$Cet&010&e & S& 2005-180T14:24 &  3.45 &  -30.30 & 1170. & 1355. & 0.01 & 0.01 & 4.30 & 4.49 & 0.84 \\
$o$Cet&012&i & S& 2005-217T01:30 &  3.45 &  -13.27 & 1108. & 1294. & 0.07 & 0.05 & 2.66 & 3.10 & 0.80 \\
$o$Cet&012&e & S& 2005-217T01:57 &  3.45 &  -13.27 & 1087. & 1274. & 0.03 & 0.02 & 3.59 & 3.74 & 0.85 \\
$\alpha$Sco&013&i & S& 2005-232T11:12 & 32.16 &  277.55 &  802. &  788. & 0.74 & 0.72 & 0.31 & 0.33 & 0.90 \\
$\alpha$Sco&013&e & S& 2005-232T14:17 & 32.16 &    8.79 &  753. &  741. & 0.19 & 0.19 & 1.65 & 1.67 & 0.90 \\
$\alpha$Tau&028&i & U& 2006-252T10:26 & 22.17 &   35.78 &  131. &  119. & 0.49 & 0.47 & 0.72 & 0.76 & 0.91 \\
$\delta$Vir&029&i & U,E& 2006-268T22:29 &  2.38 &  209.86 &  124. &  124. & 0.36 & 0.32 & 1.03 & 1.15 & 0.92 \\
$\delta$Vir&029&e & U,E& 2006-268T22:47 &  2.38 &   91.40 &  123. &  126. & 0.23 & 0.24 & 1.47 & 1.42 & 0.92 \\
$\alpha$Sco&029&i & S& 2006-269T06:35 & 32.16 &  201.05 &  717. &  735. & 0.23 & 0.23 & 1.46 & 1.49 & 0.89 \\
RLeo&030&i & S& 2006-285T01:59 &  9.55 &  335.59 &   62. &   96. & 0.40 & 0.39 & 0.90 & 0.94 & 0.89 \\
RLeo&030&e & S& 2006-285T02:51 &  9.55 &  260.51 &   60. &   94. & 0.34 & 0.35 & 1.09 & 1.04 & 0.89 \\
CWLeo&031&i & U& 2006-301T01:18 & 11.38 &  -14.16 &  210. &  384. & 0.28 & 0.27 & 1.26 & 1.32 & 0.88 \\
$\alpha$Aur&034&i & S& 2006-336T12:22 & 50.88 &   28.99 &  416. &  382. & 0.73 & 0.71 & 0.32 & 0.35 & 0.89 \\
RHya&036&i & S& 2007-001T16:27 & 29.40 &  200.10 &  332. &  407. & 0.21 & 0.18 & 1.57 & 1.71 & 0.88 \\
$\alpha$Aur&041&i & S& 2007-082T16:44 & 50.88 &   14.19 &  193. &  177. & 0.50 & 0.46 & 0.70 & 0.77 & 0.87 \\
RHya&041&i & S& 2007-088T06:01 & 29.40 & -153.47 &  104. &  143. & 0.40 & 0.36 & 0.91 & 1.02 & 0.88 \\
RHya&042&i & S& 2007-105T16:28 & 29.40 &  276.03 &  113. &  155. & 0.57 & 0.56 & 0.56 & 0.59 & 0.78 \\
$\alpha$Ori&046&i & U& 2007-163T01:57 & 11.68 &    2.46 &  726. &  686. & 0.06 & 0.05 & 2.75 & 3.05 & 0.88 \\
$\alpha$Sco&055&e & S& 2008-003T11:23 & 32.16 &   52.44 &  747. &  781. & 0.18 & 0.16 & 1.71 & 1.84 & 0.88 \\
RLeo&060&i & S& 2008-063T15:06 &  9.55 &   87.88 &  454. &  563. & 0.00 & 0.00 & 5.62 & 5.36 & 0.84 \\
RLeo&060&e & S& 2008-063T16:46 &  9.55 &  135.87 &  447. &  554. & 0.01 & 0.01 & 4.24 & 4.48 & 0.88 \\
RLeo&061&i & S& 2008-074T06:57 &  9.55 &   89.78 &  421. &  532. & 0.04 & 0.03 & 3.23 & 3.56 & 0.87 \\
RLeo&061&e & S& 2008-074T08:30 &  9.55 &  133.90 &  416. &  527. & 0.46 & 0.42 & 0.77 & 0.88 & 0.85 \\
$\alpha$TrA&063&i & S& 2008-092T02:37 & 74.19 &  275.28 &  234. &  238. & 0.82 & 0.82 & 0.20 & 0.20 & 0.93 \\
$\alpha$TrA&063&e & S& 2008-092T07:12 & 74.19 &  322.90 &  236. &  236. & 0.42 & 0.42 & 0.87 & 0.87 & 0.96 \\
RLeo&063&i & S& 2008-094T12:33 &  9.55 &   77.21 &  367. &  480. & 0.07 & 0.06 & 2.72 & 2.83 & 0.86 \\
RLeo&063&e & S& 2008-094T14:35 &  9.55 &  144.71 &  362. &  471. & 0.29 & 0.26 & 1.23 & 1.36 & 0.87 \\
RLeo&068&i & S& 2008-140T15:33 &  9.55 &   68.79 &   41. &   53. & 0.49 & 0.35 & 0.72 & 1.05 & 0.79 \\
RLeo&068&e & S& 2008-140T18:10 &  9.55 &  150.91 &   56. &   74. & 0.45 & 0.46 & 0.80 & 0.78 & ---- \\
CWLeo&070&i & S& 2008-155T13:43 & 11.38 &   69.33 &  350. &  569. & 0.43 & 0.39 & 0.84 & 0.94 & 0.89 \\
CWLeo&070&e & S& 2008-155T16:53 & 11.38 &  149.47 &  371. &  566. & 0.25 & 0.21 & 1.40 & 1.56 & 0.87 \\
\hline
\end{tabular}}}
\small

$^a$ Summation mode, S=spectrally summed, U=not spectrally summed, E=Spectrally edited.

$^b$ UTC time when star crossed 140,000 km in the ringplane.

$^c$ Elevation angle of star

$^d$ Inertial longitude of the cut at 140,000 km in the ringplane.

$^e$ Maximum Data Number after summation

$^f$ Minimum observed  transmission through the ring. Data normalized to unity in the regions 138,000-139,000 km and 141,000-142,000 km.

$^g$ Weighted average of the ratio of optical depths  
at 2.9 and 3.2 $\mu$m in the region between 139,000and 141,000 km. Only
provided where the peak optical depth exceeds 10 times the standard deviation
of the background.

\end{table}

\setcounter{table}{0}

\begin{table}
\caption{Stellar occulations of the F ring (continued)}
\label{obstab2}
\centerline{\resizebox{6in}{!}{\begin{tabular}{|l c c | c | c | c c | c c |c c|c c|c|}\hline 
Star & Rev & ingress/ & Sum$^a$ & Date$^b$ & B$^c$ & Longitude$^d$ & Max DN$^e$ & Max DN$^e$ & Min T & Min T & Max $\tau$ & Max $\tau$ & $\langle\rho\rangle$ \\ 
 & & egress & & & (deg) & (deg) & 2.92 $\mu$m & 3.19$\mu$m & 2.92 $\mu$m & 3.19$\mu$m & 
 2.92 $\mu$m & 3.19$\mu$m &\\
 \hline
$\gamma$Cru&071&i & S& 2008-160T08:13 & 62.35 &  187.65 &  477. &  507. & 0.77 & 0.77 & 0.26 & 0.26 & ---- \\
CWLeo&071&i & S& 2008-162T17:13 & 11.38 &   70.65 &  174. &  365. & 0.20 & 0.19 & 1.59 & 1.67 & 0.94 \\
$\gamma$Cru&072&i & S& 2008-167T11:30 & 62.35 & -172.71 &  641. &  673. & 0.78 & 0.76 & 0.25 & 0.27 & 0.86 \\
$\gamma$Cru&073&i & S& 2008-174T14:39 & 62.35 &  186.98 &  654. &  685. & 0.55 & 0.52 & 0.60 & 0.65 & 0.90 \\
CWLeo&074&i & S& 2008-184T00:16 & 11.38 &   73.85 &  140. &  230. & 0.16 & 0.13 & 1.82 & 2.04 & 0.87 \\
CWLeo&074&e & S& 2008-184T03:09 & 11.38 &  144.75 &  142. &  233. & 0.34 & 0.30 & 1.09 & 1.22 & 0.87 \\
RLeo&075&i & S& 2008-191T04:10 &  9.55 &   67.95 &  270. &  397. & 0.61 & 0.58 & 0.49 & 0.55 & 0.84 \\
RLeo&075&e & S& 2008-191T06:59 &  9.55 &  149.97 &  270. &  396. & 0.07 & 0.05 & 2.65 & 2.96 & 0.90 \\
$\gamma$Cru&077&i & S& 2008-202T18:16 & 62.35 &  186.19 &  139. &  153. & 0.28 & 0.28 & 1.26 & 1.26 & 0.95 \\
RLeo&077&i & S& 2008-205T06:22 &  9.55 &   70.60 &  288. &  418. & 0.11 & 0.10 & 2.23 & 2.31 & 0.87 \\
RLeo&077&e & S& 2008-205T09:03 &  9.55 &  147.28 &  289. &  418. & 0.35 & 0.31 & 1.04 & 1.17 & 0.85 \\
$\gamma$Cru&078&i & S& 2008-209T19:17 & 62.35 &  186.01 &  280. &  286. & 0.69 & 0.65 & 0.38 & 0.43 & 0.88 \\
$\eta$Car&078&e & S& 2008-210T03:29 & 62.47 &  336.59 &   65. &   90. & 0.31 & 0.34 & 1.16 & 1.07 & 0.98 \\
$\beta$Gru&078&i & S& 2008-210T09:15 & 43.38 & -103.88 &  271. &  297. & 0.93 & 0.93 & 0.07 & 0.07 & ---- \\
$\gamma$Cru&079&i & S& 2008-216T11:55 & 62.35 &  185.09 &  697. &  726. & 0.84 & 0.81 & 0.18 & 0.21 & 0.88 \\
RSCnc&080&i & S& 2008-226T01:14 & 29.96 &   50.08 &  309. &  361. & 0.32 & 0.31 & 1.14 & 1.18 & 0.90 \\
RSCnc&080&e & S& 2008-226T08:18 & 29.96 &  161.65 &  295. &  346. & 0.44 & 0.41 & 0.82 & 0.90 & 0.90 \\
$\gamma$Cru&081&i & S& 2008-231T06:03 & 62.35 &  184.53 &  590. &  611. & 0.95 & 0.94 & 0.05 & 0.06 & 0.81 \\
$\gamma$Cru&082&i & S& 2008-238T14:39 & 62.35 &  184.24 &  719. &  742. & 0.83 & 0.81 & 0.19 & 0.21 & 0.88 \\
RSCnc&085&i & S& 2008-262T21:39 & 29.96 &   51.47 &  310. &  369. & 0.27 & 0.25 & 1.33 & 1.40 & 0.91 \\
RSCnc&085&e & S& 2008-263T04:37 & 29.96 &  159.88 &  309. &  367. & 0.66 & 0.63 & 0.42 & 0.46 & 0.94 \\
$\gamma$Cru&086&i & S& 2008-268T02:19 & 62.35 &  183.55 & 1030. & 1081. & 0.84 & 0.83 & 0.17 & 0.19 & 0.89 \\
RLeo&086&i & S& 2008-271T10:01 &  9.55 &   86.47 &  654. &  906. & 0.87 & 0.85 & 0.14 & 0.16 & 0.84 \\
RLeo&086&e & S& 2008-271T11:35 &  9.55 &  132.01 &  650. &  901. & 0.69 & 0.67 & 0.37 & 0.40 & 0.86 \\
RSCnc&087&i & S& 2008-277T15:26 & 29.96 &   52.12 &  326. &  383. & 0.17 & 0.12 & 1.78 & 2.14 & 0.84 \\
RSCnc&087&e & S& 2008-277T22:20 & 29.96 &  159.11 &  322. &  393. & 0.25 & 0.25 & 1.40 & 1.37 & 0.98 \\
RLeo&087&i & U& 2008-278T18:51 &  9.55 &   87.43 &  304. &  417. & 0.84 & 0.81 & 0.18 & 0.21 & 0.85 \\
RLeo&087&e & U& 2008-278T20:23 &  9.55 &  131.05 &  303. &  416. & 0.05 & 0.05 & 2.95 & 3.02 & 0.93 \\
$\gamma$Cru&089&i & S& 2008-290T03:30 & 62.35 &  183.34 &  704. &  733. & 0.77 & 0.75 & 0.27 & 0.29 & 0.89 \\
RSCnc&092&i & S& 2008-315T00:42 & 29.96 &   69.72 &  223. &  263. & 0.10 & 0.08 & 2.34 & 2.51 & 0.96 \\
$\gamma$Cru&093&i & S& 2008-320T15:30 & 62.35 & -157.95 &  466. &  499. & 0.54 & 0.53 & 0.62 & 0.63 & 0.94 \\
$\gamma$Cru&094&i & S& 2008-328T00:26 & 62.35 &  192.12 &  269. &  283. & 0.09 & 0.10 & 2.37 & 2.25 & 0.89 \\
$\epsilon$Mus&094&i & S& 2008-328T06:47 & 72.77 &  245.59 &  190. &  209. & 0.55 & 0.52 & 0.59 & 0.65 & 0.83 \\
$\gamma$Cru&096&i & S& 2008-343T10:54 & 62.35 & -171.80 &  209. &  227. & 0.49 & 0.48 & 0.70 & 0.73 & 0.89 \\
$\gamma$Cru&097&i & S& 2008-351T10:12 & 62.35 &  188.11 &  856. &  925. & 0.37 & 0.34 & 1.00 & 1.08 & 0.88 \\
$\gamma$Cru&100&i & S& 2009-012T09:24 & 62.35 & -149.72 &  390. &  403. & 0.79 & 0.77 & 0.24 & 0.26 & 0.86 \\
$\alpha$TrA&100&i & S& 2009-013T02:58 & 74.19 &  237.01 &   25. &   19. & 0.16 & 0.40 & 1.81 & 0.91 & ---- \\
$\alpha$TrA&100&e & S& 2009-013T10:44 & 74.19 &  334.80 &   20. &   11. & 0.35 & 0.12 & 1.04 & 2.13 & ---- \\
$\gamma$Cru&101&i & S& 2009-021T23:10 & 62.35 &  210.27 &  418. &  438. & 0.60 & 0.57 & 0.52 & 0.56 & 0.89 \\
$\gamma$Cru&102&i & S& 2009-031T12:23 & 62.35 & -149.92 &  993. & 1047. & 0.79 & 0.76 & 0.24 & 0.27 & 0.87 \\
TXCam&102&i & S& 2009-034T23:01 & 61.29 &  341.45 &   49. &   69. & 0.61 & 0.63 & 0.50 & 0.46 & 1.01 \\
$\gamma$Cru&104&i & S& 2009-053T08:17 & 62.35 &  255.06 &  220. &  238. & 0.51 & 0.48 & 0.68 & 0.73 & 0.96 \\
$\beta$Peg&104&i & S& 2009-057T08:33 & 31.68 &   -3.47 &  293. &  309. & 0.91 & 0.89 & 0.10 & 0.12 & 0.85 \\
$\gamma$Cru&106&i & S& 2009-077T06:43 & 62.35 &  254.88 &  803. &  834. & 0.75 & 0.73 & 0.29 & 0.32 & 0.89 \\
$\gamma$Cru&106&e & S& 2009-077T12:47 & 62.35 &  306.80 &  757. &  815. & 0.38 & 0.38 & 0.96 & 0.96 & 0.93 \\
RCas&106&i & S& 2009-081T20:43 & 56.04 &   44.25 &  128. &  169. & 0.45 & 0.42 & 0.79 & 0.86 & 0.90 \\
$\beta$Peg&108&i & S& 2009-095T13:53 & 31.68 &    5.32 &  289. &  306. & 0.69 & 0.66 & 0.37 & 0.42 & 0.88 \\
$\alpha$Aur&110&i & S& 2009-129T10:35 & 50.88 &  -34.08 &  319. &  291. & 0.38 & 0.35 & 0.96 & 1.04 & 0.89 \\
$\alpha$Aur&110&e & S& 2009-129T17:59 & 50.88 & -127.08 &  257. &  258. & 0.69 & 0.67 & 0.37 & 0.40 & 0.88 \\
$\alpha$Sco&115&i & S& 2009-208T22:14 & 32.16 &  173.78 & 2705. & 2866. & 0.12 & 0.10 & 2.11 & 2.31 & 0.90 \\
$\alpha$Ori&117&i & S& 2009-239T07:26 & 11.68 &   31.20 &  513. &  542. & 0.21 & 0.19 & 1.54 & 1.68 & 0.93 \\
\hline
\end{tabular}}}
\end{table}

\begin{table}
\caption{Stellar occultations of the Encke Gap ringlets}
\label{egaptab}
\centerline{\resizebox{6in}{!}{\begin{tabular}{|l c c | c | c | c c c| c c |ccc|ccc|ccc|}\hline 
Star & Rev & ingress/ & Sum& Date & B& Longitude  & Longitude & Max DN   & Max DN & \multicolumn{3}{|c|}{Inner Ringlet} & \multicolumn{3}{|c|}{Central Ringlet} & \multicolumn{3}{|c|}{Outer Ringlet}  \\
& & egress & & &  & of ring & of Pan & & & Max $\tau$ & Max $\tau$ & $<{\rho}>^a$ & Max $\tau$ & Max $\tau$ & $\langle\rho\rangle^b$ & Max $\tau$ & Max $\tau$ & $<{\rho}>^c$ \\ 
 & &  & & & (deg) & (deg) & (deg) & 2.92 $\mu$m & 3.19$\mu$m & 2.92 $\mu$m & 3.19$\mu$m & & 
 2.92 $\mu$m & 3.19$\mu$m & & 2.92 $\mu$m & 3.19$\mu$m & \\
\hline
$o$Cet&008&i & S& 2005-144T05:14 &  3.45 &   16.72 & -106.73 &  999. & 1192. & 0.0344 & 0.0333 & ---- & 0.0344 & 0.0343 & ---- & 0.1154 & 0.1254 & 0.96 \\
$o$Cet&008&e & S& 2005-144T07:50 &  3.45 &  -43.68 &  -38.83 & 1001. & 1188. & 0.1217 & 0.1199 & 0.98 & 0.0169 & 0.0171 & ---- & 0.0180 & 0.0196 & 1.08 \\
$o$Cet&009&i & S& 2005-162T08:24 &  3.45 &    5.38 &   84.64 &  993. & 1164. & ---- & ---- & ---- & 0.0668 & 0.0697 & 0.98 & ---- & ---- & ---- \\
$o$Cet&009&e & S& 2005-162T10:06 &  3.45 &  -32.19 &  128.65 & 1005. & 1181. & 0.0214 & 0.0249 & ---- & 0.0435 & 0.0416 & ---- & 0.0238 & 0.0254 & ---- \\
$o$Cet&010&i & S& 2005-180T13:09 &  3.45 &   -6.38 &  -43.17 & 1176. & 1363. & 0.2788 & 0.2894 & 0.96 & 4.7709 & 5.2434 & 0.94 & 0.0527 & 0.0544 & 0.99 \\
$o$Cet&010&e & S& 2005-180T13:55 &  3.45 &  -20.34 &  -23.14 & 1187. & 1376. & 0.0417 & 0.0403 & 0.95 & 0.1973 & 0.2091 & 0.94 & 0.0566 & 0.0551 & 1.00 \\
$\delta$Vir&029&i & U,E& 2006-268T22:29 &  2.38 &  208.57 &  112.96 &  123. &  125. & 2.0838 & 1.9774 & 0.97 & ---- & ---- & ---- & ---- & ---- & ---- \\
RHya&042&i & S& 2007-105T17:24 & 29.40 &  279.35 &   78.51 &  113. &  157. & ---- & ---- & ---- & ---- & ---- & ---- & 0.2815 & 0.3127 & 0.95 \\
RLeo&060&e & S& 2008-063T16:32 &  9.55 &  130.43 &  -55.79 &  460. &  584. & ---- & ---- & ---- & 0.0241 & 0.0246 & ---- & ---- & ---- & ---- \\
RLeo&061&i & S& 2008-074T07:12 &  9.55 &   95.94 &  107.29 &  432. &  553. & 0.0626 & 0.0561 & 1.02 & ---- & ---- & ---- & ---- & ---- & ---- \\
RLeo&061&e & S& 2008-074T08:15 &  9.55 &  127.74 &  134.56 &  431. &  557. & 0.0361 & 0.0342 & ---- & ---- & ---- & ---- & ---- & ---- & ---- \\
RLeo&063&i & S& 2008-094T12:42 &  9.55 &   80.59 &  171.27 &  369. &  496. & ---- & ---- & ---- & 0.0259 & 0.0297 & ---- & ---- & ---- & ---- \\
CWLeo&074&i & S& 2008-184T00:28 & 11.38 &   77.01 &   34.87 &  146. &  248. & ---- & ---- & ---- & 0.5296 & 0.5531 & 1.00 & ---- & ---- & ---- \\
CWLeo&074&e & S& 2008-184T02:57 & 11.38 &  141.60 &   99.84 &  156. &  264. & ---- & ---- & ---- & 0.3642 & 0.3706 & 0.98 & ---- & ---- & ---- \\
RLeo&077&i & S& 2008-205T06:32 &  9.55 &   73.42 &   19.67 &  298. &  440. & ---- & ---- & ---- & 0.1009 & 0.1053 & 0.95 & ---- & ---- & ---- \\
RLeo&077&e & S& 2008-205T08:53 &  9.55 &  144.46 &   81.33 &  300. &  443. & ---- & ---- & ---- & 0.0401 & 0.0334 & ---- & 0.0612 & 0.0683 & 1.03 \\
RSCnc&080&i & S& 2008-226T01:28 & 29.96 &   51.58 &   74.51 &  311. &  363. & ---- & ---- & ---- & ---- & ---- & ---- & 0.2367 & 0.2459 & 0.93 \\
RSCnc&085&e & S& 2008-263T04:22 & 29.96 &  158.29 &  -86.66 &  314. &  377. & 0.0942 & 0.1015 & 1.00 & ---- & ---- & ---- & ---- & ---- & ---- \\
RLeo&086&i & S& 2008-271T10:15 &  9.55 &   92.33 &   34.91 &  668. &  932. & ---- & ---- & ---- & 0.0742 & 0.0730 & 0.98 & 0.0654 & 0.0668 & 0.97 \\
RSCnc&087&e & S& 2008-277T22:05 & 29.96 &  157.48 &  139.90 &  327. &  400. & ---- & ---- & ---- & 0.0325 & 0.0355 & ---- & ---- & ---- & ---- \\
RLeo&087&e & U& 2008-278T20:07 &  9.55 &  124.79 &   -5.27 &  315. &  439. & 0.0735 & 0.0643 & ---- & ---- & ---- & ---- & ---- & ---- & ---- \\
$\gamma$Cru&104&e & S& 2009-053T13:36 & 62.35 &  301.71 & -104.92 &  167. &  184. & ---- & ---- & ---- & 0.1059 & 0.1130 & ---- & ---- & ---- & ---- \\
$\beta$Peg&104&i & S& 2009-057T08:42 & 31.68 &   -4.18 &  111.21 &  297. &  314. & ---- & ---- & ---- & ---- & ---- & ---- & 0.1348 & 0.1384 & 0.92 \\
$\alpha$Ori&117&i & S& 2009-239T07:32 & 11.68 &   31.82 & -101.37 &  525. &  552. & 2.1787 & 2.2825 & 0.97 & ---- & ---- & ---- & ---- & ---- & ---- \\
\hline
\end{tabular}}}

Data normalized to unity in the ranges 133,510-133,540 and 133,650-133,700 km.

$^a$ Weighted average of the ratio of optical depths  at 2.9 and 3.2 $\mu$m
in the range between 133,450 and 133,510 km. Only
provided where the peak optical depth exceeds 10 times the standard deviation
of the background.

$^b$ Weighted average of the ratio of optical depths  at 2.9 and 3.2 $\mu$m
in the range between 133,540 and 133,650 km. Only
provided where the peak optical depth exceeds 10 times the standard deviation
of the background.

$^c$ Weighted average of the ratio of optical depths  at 2.9 and 3.2 $\mu$m
in the range between 133,700 and 133,730 km. Only
provided where the peak optical depth exceeds 10 times the standard deviation
of the background.

\end{table}

\begin{table}
\caption{Stellar occulations of the Charming Ringlet}
\label{charmtab}
\centerline{\resizebox{6in}{!}{\begin{tabular}{|l c c | c | c | c c c |c c|c c|c|}\hline 
Star & Rev & ingress/ & Sum& Date & B& Longitude  & Long. of Sun & Max DN & Max DN & Max $\tau\sin B^a$ & Max $\tau \sin B^a$ & $<{\rho}>^b$ \\ 
 & & egress & & & (deg) & (deg) & (deg) & 2.92 $\mu$m & 3.19$\mu$m & 2.92 $\mu$m & 3.19$\mu$m  
 &\\
 \hline
 $o$Cet&008&i & S& 2005-144T05:53 &  3.45 &    2.37 &  172.68 & 1000. & 1185. & 0.0033 & 0.0038 & 0.83 \\
$o$Cet&008&e & S& 2005-144T07:11 &  3.45 &  -29.33 &  172.68 & 1007. & 1196. & 0.0024 & 0.0026 & 0.83 \\
$\alpha$Ori&026&i & S& 2006-204T16:45 & 11.68 &   -2.26 & -171.67 &  967. & 1013. & 0.0039 & 0.0047 & 0.87 \\
$\alpha$Sco&029&i & S& 2006-269T07:37 & 32.16 &  194.22 & -169.40 &  713. &  729. & 0.0073 & 0.0091 & ---- \\
RLeo&030&e & S& 2006-285T02:40 &  9.55 &  274.44 & -168.85 &   64. &  104. & 0.0279 & 0.0251 & ---- \\
CWLeo&031&i & U& 2006-301T01:26 & 11.38 &  -22.60 & -168.29 &  164. &  319. & 0.0122 & 0.0090 & ---- \\
$\alpha$Sco&055&e & S& 2008-003T09:50 & 32.16 &   64.25 & -153.79 &  728. &  761. & 0.0097 & 0.0108 & ---- \\
RLeo&063&i & S& 2008-094T13:08 &  9.55 &   94.78 & -150.86 &  371. &  486. & 0.0063 & 0.0065 & ---- \\
RLeo&063&e & S& 2008-094T14:01 &  9.55 &  127.14 & -150.86 &  372. &  491. & 0.0061 & 0.0068 & ---- \\
RLeo&075&i & S& 2008-191T04:41 &  9.55 &   79.61 & -147.78 &  279. &  408. & 0.0067 & 0.0066 & ---- \\
RLeo&075&e & S& 2008-191T06:28 &  9.55 &  138.31 & -147.78 &  280. &  411. & 0.0072 & 0.0075 & ---- \\
RLeo&077&i & S& 2008-205T06:56 &  9.55 &   83.90 & -147.34 &  296. &  429. & 0.0053 & 0.0063 & ---- \\
RLeo&077&e & S& 2008-205T08:29 &  9.55 &  133.98 & -147.34 &  297. &  433. & 0.0097 & 0.0073 & ---- \\
$\gamma$Cru&097&i & S& 2008-351T11:00 & 62.35 &  187.44 & -142.76 &  808. &  869. & 0.0140 & 0.0161 & ---- \\
$\alpha$Sco&115&i & S& 2009-208T23:53 & 32.16 &  170.44 & -135.88 & 2719. & 2877. & 0.0062 & 0.0064 & ---- \\
$\alpha$Ori&117&i & S& 2009-239T07:45 & 11.68 &   33.75 & -134.95 &  547. &  572. & 0.0064 & 0.0064 & ---- \\
 \hline
\end{tabular}}}

$^a$ Normal optical depth, data normalized to unity between
119,980 and 120,020 km.

$^b$ Weighted average of the ratio of optical depths  at 2.9 and 3.2 $\mu$m
in the range between 119,880 and 119,980 km. Only
provided where the peak optical depth exceeds 10 times the standard deviation
of the background.

\end{table}

\begin{table}
\caption{Features with distinctive transmission spectra in the F ring}
\label{clumptab}
\resizebox{6in}{!}{\begin{tabular}{|c|c|c|c|c|c|c|c|c|c|c|c|c|c|}\hline
Star & Rev & Ingress/ & SCET (seconds) & Subspacecraft & Subspacecraft & Radius & Longitude & S & Peak $\tau_b$ & $\tau_d^a$ & Predicted & Prometheus & Prometheus \\
 & & Egress & & Longitude & Latitude & (km) & (deg) & & & &  Radius (km) & Radius (km) & Long. (km) \\ \hline 
$\alpha$Sco & 013 & e & 1503240353.106 &  -26.9 &  32.2 &   140546.2 &    9.99 &  6.5 & 1.47 &  0.07 &   140530.7 &   139067.8 &  -94.27 \\
$\alpha$Sco & 029 & i & 1537945646.520 &  -82.1 &  32.2 &   139916.0 & -158.36 &  7.7 & 1.24 &  0.47 &   139906.8 &   139250.6 &    5.88 \\
$\alpha$Sco & 055 & e & 1578052761.191 &  -44.5 &  32.2 &   140008.4 &   51.46 &  5.7 & 0.86 &  0.97 &   140013.2 &   139132.0 &  102.30 \\
$\gamma$Cru & 073 & i & 1592839004.290 & -144.2 &  62.3 &   140182.9 & -172.91 &  3.3 & 0.24 &  0.31 &   140193.6 &   139347.2 &  167.64 \\
$\gamma$Cru & 077 & i & 1595271275.846 & -144.6 &  62.3 &   139906.5 & -173.71 &  3.1 & 1.50 & -0.41 &   139901.6 &   139686.0 &  140.67 \\
 RSCnc & 087 & e & 1601766034.355 & -174.4 & -30.0 &   140478.2 &  159.87 &  4.6 & 1.61 &  -0.03 &   140492.0 &   139685.2 &    7.00 \\
$\gamma$Cru & 093 & i & 1605456599.050 & -137.4 &  62.3 &   140078.4 & -158.08 &  6.6 & 0.66 &  0.06 &   140090.0 &   139147.5 & -107.59 \\
$\gamma$Cru & 094 & i & 1606093583.558 & -140.0 &  62.3 &   139947.7 & -167.88 &  3.1 & 2.53$^b$ &  0.07 &   139959.8 &   139188.8 &  -97.80 \\
$\gamma$Cru & 097 & i & 1608115920.990 & -141.6 &  62.3 &   139954.1 & -171.83 &  3.6 & 0.33 &  0.65 &   139945.0 &   139182.2 &  -31.62 \\
$\gamma$Cru & 104 & i & 1613984114.746 & -124.3 &  62.3 &   140511.1 & -107.02 &  3.9 & 0.58 &  0.06 &   140510.8 &   139174.6 & -103.71 \\
$\gamma$Cru & 106 & e & 1616074308.867 & -114.5 &  62.3 &   140523.3 &  -51.12 &  3.4 & 0.18 &  0.13 &   140531.9 &   139648.6 &   63.74 \\
$\gamma$Cru & 106 & e & 1616074311.376 & -114.5 &  62.3 &   140530.2 &  -51.12 & 10.9 & 1.06 &  0.25 &   140531.9 &   139648.6 &   63.73 \\
$\alpha$Sco & 115 & i & 1627426537.697 &  -70.3 &  32.2 &   140078.1 &  174.10 & 13.5 & 0.74 &  0.66 &   140110.9 &   139370.7 & -172.84 \\
$\alpha$Sco & 115 & i & 1627426568.854 &  -70.2 &  32.2 &   139972.3 &  174.09 &  5.1 & 0.11 &  0.09 &   140110.8 &   139369.6 & -172.63 \\
\hline
\end{tabular}}
$^a$ Note a low-pass filter is applied to the $\tau_d$ profile prior to calculating these estimates, so 
the values in this table are somewhat different from those shown in Figures~\ref{clumpplot}-~\ref{clumpplot2}.

$^b$ the peak calculated $\tau_b$ is 4.33, but the peak $\tau_{3.2}$ is 2.53, and this a more reliable
estimate of the optical depth of this feature.
\end{table}

\end{document}